\documentclass[fleqn,usenatbib]{mnras}

\usepackage{newtxtext,newtxmath}

\usepackage[T1]{fontenc}

\DeclareRobustCommand{\VAN}[3]{#2}
\let\VANthebibliography\thebibliography
\def\thebibliography{\DeclareRobustCommand{\VAN}[3]{##3}\VANthebibliography}

\usepackage{graphicx}	
\usepackage{amsmath}	
\usepackage{braket}
\usepackage{mathtools}

\usepackage{multirow}
\usepackage{rotating}
\usepackage{tabularx}
\usepackage{subcaption}
\usepackage{outlines}
\usepackage{booktabs}
\usepackage{natbib}
\newcolumntype{s}{>{\hsize=.3\hsize}X}
\usepackage{orcidlink}
\usepackage{longtable}

\title[Eccentricity distribution of giant planets]{Investigating the eccentricity distribution of transiting, long-period giant planets}

\author[A. Alqasim et al.]{
Ahlam Alqasim\textsuperscript{\hyperlink{inst:1}{1},\hyperlink{inst:2}{2}}\thanks{E-mail: ahlam.alqasim.17@ucl.ac.uk}\orcidlink{0000-0001-5102-5505},
Teruyuki Hirano\textsuperscript{\hyperlink{inst:2}{2},\hyperlink{inst:3}{3}}\orcidlink{0000-0003-3618-7535}, 
Yasunori Hori\textsuperscript{\hyperlink{inst:2}{2},\hyperlink{inst:3}{3}}\orcidlink{0000-0003-4676-0251}, 
Daisuke Kawata\textsuperscript{\hyperlink{inst:1}{1},\hyperlink{inst:3}{3}}\orcidlink{0000-0001-8993-101X}, 
John Livingston\textsuperscript{\hyperlink{inst:2}{2},\hyperlink{inst:3}{3},\hyperlink{inst:4}{4}}\orcidlink{0000-0002-4881-3620}, 
\newauthor Steve B. Howell\textsuperscript{\hyperlink{inst:5}{5}}\orcidlink{0000-0002-2532-2853}, 
Antonino F. Lanza\textsuperscript{\hyperlink{inst:6}{6}}\orcidlink{0000-0001-5928-7251}, 
Andrew W. Mann\textsuperscript{\hyperlink{inst:7}{7}}\orcidlink{0000-0003-3654-1602}, 
Carl Ziegler\textsuperscript{\hyperlink{inst:8}{8}}\orcidlink{0000-0002-0619-7639}, 
C\'{e}sar Brice\~{n}o\textsuperscript{\hyperlink{inst:9}{9}}, 
\newauthor Charles A. Beichman\textsuperscript{\hyperlink{inst:10}{10}}\orcidlink{0000-0002-5627-5471}, 
David R. Ciardi\textsuperscript{\hyperlink{inst:11}{11}}\orcidlink{0000-0002-5741-3047}, 
Ivan A. Strakhov\textsuperscript{\hyperlink{inst:12}{12}}\orcidlink{0000-0003-0647-6133}, 
Michael B. Lund\textsuperscript{\hyperlink{inst:11}{11}}\orcidlink{0000-0003-2527-1598}, 
and Nicholas Law\textsuperscript{\hyperlink{inst:7}{7}} 
\\
\textsuperscript{\hypertarget{inst:1}{1}}UCL Mullard Space Science Laboratory, Holmbury Hill Road, Dorking, Surrey, RH5 6NT, UK \\
\textsuperscript{\hypertarget{inst:2}{2}}Astrobiology Center, 2-21-1 Osawa, Mitaka, Tokyo 181-8588, Japan \\
\textsuperscript{\hypertarget{inst:3}{3}}National Astronomical Observatory of Japan, 2-21-1 Osawa, Mitaka, Tokyo 181-8588, Japan \\
\textsuperscript{\hypertarget{inst:4}{4}}Department of Astronomical Science, The Graduate University for Advanced Studies, 2-21-1 Osawa, Mitaka, Tokyo, 181-8588, Japan \\
\textsuperscript{\hypertarget{inst:5}{5}}NASA Ames Research Center, Moffett Field, CA 94035, USA \\
\textsuperscript{\hypertarget{inst:6}{6}}INAF-Osservatorio Astrofisico di Catania, Via S. Sofia, 78 - 95123 Catania, Italy \\
\textsuperscript{\hypertarget{inst:7}{7}}Department of Physics and Astronomy, The University of North Carolina at Chapel Hill, Chapel Hill, NC 27599-3255, USA \\
\textsuperscript{\hypertarget{inst:8}{8}}Department of Physics, Engineering and Astronomy, Stephen F. Austin State University, 1936 North St, Nacogdoches, TX 75962, USA \\
\textsuperscript{\hypertarget{inst:9}{9}}Cerro Tololo Inter-American Observatory, Casilla 603, La Serena, Chile \\
\textsuperscript{\hypertarget{inst:10}{10}}NASA Exoplanet Science Institute-Caltech/IPAC and Jet Propulsion Laboratory, Pasadena, CA 91125, USA \\
\textsuperscript{\hypertarget{inst:11}{11}}NASA Exoplanet Science Institute-Caltech/IPAC, 1200 E. California Blvd, Pasadena, CA 91125 USA \\
\textsuperscript{\hypertarget{inst:12}{12}}Sternberg Astronomical Institute, Moscow State University, 119992, Universitetski pr., 13, Moscow, Russia
}

\date{Accepted XXX. Received YYY; in original form ZZZ}

\pubyear{\the\year{}}

\begin{document}
\label{firstpage}
\pagerange{\pageref{firstpage}--\pageref{lastpage}}
\maketitle

\begin{abstract}
Eccentric giant planets are predicted to have acquired their eccentricity through two major mechanisms: the Kozai-Lidov effect or planet-planet scattering, but it is normally difficult to separate the two mechanisms and determine the true eccentricity origin for a given system.
In this work, we focus on a sample of 92 transiting, long-period giant planets (TLGs) as part of an eccentricity distribution study for this planet population in order to understand their eccentricity origin. 
Using archival high-contrast imaging observations, public stellar catalogs, precise \textit{Gaia} astrometry, and the NASA Exoplanet Archive database, we explored the eccentricity distribution correlation with different planet and host-star properties of our sample.
We also homogeneously characterized the basic stellar properties for all 86 host-stars in our sample, including stellar age and metallicity.
We found a correlation between eccentricity and stellar metallicity, where lower-metallicity stars ([Fe/H] $\leq$ 0.1) did not host any planets beyond $e > 0.4$, while higher-metallicity stars hosted planets across the entire eccentricity range.
Interestingly, we found no correlation between the eccentricity distribution and the presence of stellar companions, indicating that planet-planet scattering is likely a more dominant mechanism than the Kozai-Lidov effect for TLGs.
This is further supported by an anti-correlation trend found between planet multiplicity and eccentricity, as well as a lack of strong tidal dissipation effects for planets in our sample, which favor planet-planet scattering scenarios for the eccentricity origin.
\end{abstract}




\section{Introduction}\label{sec:intro}

Giant exoplanets have been suggested to have a significant effect on the formation and evolution of planetary systems \citep{levison2003giantsrole, childs2019giantseffect}.
Eccentric giant planets could have acquired their eccentricity through two major mechanisms: the Kozai-Lidov effect \citep{fabrycky2007kozai} or planet-planet scattering \citep{naoz2011planetplanet}.
However, it is often difficult to differentiate between the two mechanisms for any single system \citep{juric2008dynamical}.
Even methods such as the Rossiter–McLaughlin effect do not help much in distinguishing between these two scenarios \citep{beauge2012scattering}.
Up to a few $M_{\text{Jup}}$, eccentric planets are expected to have evolved via planet-planet scattering \citep{bitsch2020eccdist}.
Additionally, planet scattering followed by interactions with outer planets can also excite planets to high eccentricities \citep{nagasawa2011orbital}. 
In contrast, \cite{nagasawa2008formation} found that the Kozai-Lidov mechanism in outer planets can cause the formation and eccentricity excitation of close-in planets.
This is also in agreement with the findings of \cite{bonomo2017gaps} for transiting hot Jupiters, who found that those planets are consistent with formation through high-eccentricity migration.
Thus, inner planets with moderate eccentricities could have evolved via this mechanism without the presence of any close-in companions.
Giant planets have been predicted to be more likely found in multi-planet systems \citep{bitsch2020eccdist}. 
\cite{ida2013formation} showed that distant giant companions could be formed with nearly circular orbits via scattered residual cores from emerging gas giants.
A population study on a statistical level for the observed distribution of such planets could provide better insights into their eccentricity origins and evolution history.

Eccentricity distributions of transiting close-in giant planets (namely, Hot Jupiters) have been extensively studied \citep{knutson2014friendshj, bonomo2017gaps}, with many follow-up campaigns to try to search for predicted long-period companions \citep{ngo2016friendshj}. 
However, the eccentricity distributions of more distant giant planets (e.g. warmer and longer-period Jupiter, Saturn, and Neptune-like exoplanets) have not been well-studied, and their eccentricity and formation pathways tend to differ from Hot Jupiters, which makes them all the more important to explore.
Thus, investigating the planet eccentricity correlations of distant giant planets on a statistical level would enable us to validate whether different theoretical predictions reflect the observed planet distribution and their system properties. 
It would also enable us to probe which mechanism could be responsible for the eccentricity origin of planets in our sample.
The analysis presented in this paper is distinct and new in comparison to previous works, given that we are only focusing on transiting, long-period giant planets.

Our motivation is to (1) investigate the correlation of different planet and stellar properties with the eccentricity distribution for our sample of long-period exoplanets, and (2) try to distinguish between different evolution scenarios (e.g. Kozai-Lidov effect, planet-planet scattering, etc.) that could have caused planets in our sample to become eccentric.
Our population study utilizes archival high-contrast imaging observations, public stellar catalogs, precise \textit{Gaia} astrometry, and the NASA Exoplanet Archive database, to enable us to better understand and probe the origin of the eccentricity for our target sample.

Our paper is structured as follows: in Section \ref{sec:sample}, we describe the criteria used to select our sample. 
Section \ref{sec:jaxstar} describes the homogeneous characterization of the basic stellar properties for our host-stars. 
In Section \ref{sec:internal}, we describe the internal composition modeling performed for the planets in our sample.
Section \ref{sec:eccdist} details how we construct our eccentricity distributions and measure their significance in comparison with different properties.
We present our results and discuss the implications of our findings in Section \ref{sec:results_discussion}, and any possible biases of the study in Section \ref{sec:biases}.
We end with the summary and conclusions in Section \ref{sec:summary}.

\section{Sample Selection}\label{sec:sample}

The focus of our investigation is on transiting, long-period giant planets (henceforth referred to as TLGs), which are less vulnerable to tidal circularization.
As such, we selected a planet sample with a minimum mass of 10 $M{_{\oplus}}$ and with orbital periods of $P$ > 10 days. 
We place this minimum mass constraint to avoid biases on the lower eccentricity, since the eccentricity measurements are not as reliable below 10 $M{_{\oplus}}$ and are often set to 0.
We also placed the requirement that planets in our sample are transiting and have been observed with radial velocity instruments.
We limited our sample to transiting systems to ensure that we have a radius measurement of the planet, enabling us to utilize the planet density in our study when combined with the mass measurements from radial velocities.
We required that all planets in our sample had reported error measurements for radius, mass and eccentricity to ensure the reliability of the results, in particular with regards to eccentricity.

We extracted our target sample from the NASA Exoplanet Archive\footnote{\url{https://exoplanetarchive.ipac.caltech.edu}} \citep{akeson2013nexi} (henceforth referred to as NEXA), as of 9 May 2024.
We use the Planetary Systems Composite Parameters (PSCompPars) table, which provides a more statistical view of the known exoplanet population and their host environments according to NEXA.
We queried the table using NEXA's TAP service\footnote{\url{https://exoplanetarchive.ipac.caltech.edu/docs/TAP/usingTAP.html}}.
We summarize the criteria used for our sample selection as follows:
\begin{enumerate}
    \item Error requirement: The error measurements of the planet radius, mass, period, and eccentricity are available (non-null). This is filtered using the upper error bound columns.
    \item Mass cut: $M{_{\text{p}}}$ > 10 $M{_{\oplus}}$
    \item Period cut: $P$ > 10 days
    \item Transit flag: \texttt{tran\_flag} = 1
    \item Radial velocity flag: \texttt{rv\_flag} = 1
    \item Metallicity flag: \texttt{st\_metratio} = [Fe/H]
    \item Discovery method: \texttt{discoverymethod} = `Radial Velocity' or `Transit'
\end{enumerate}

\begin{figure*}
  \centering
  \includegraphics[width=\linewidth]
  {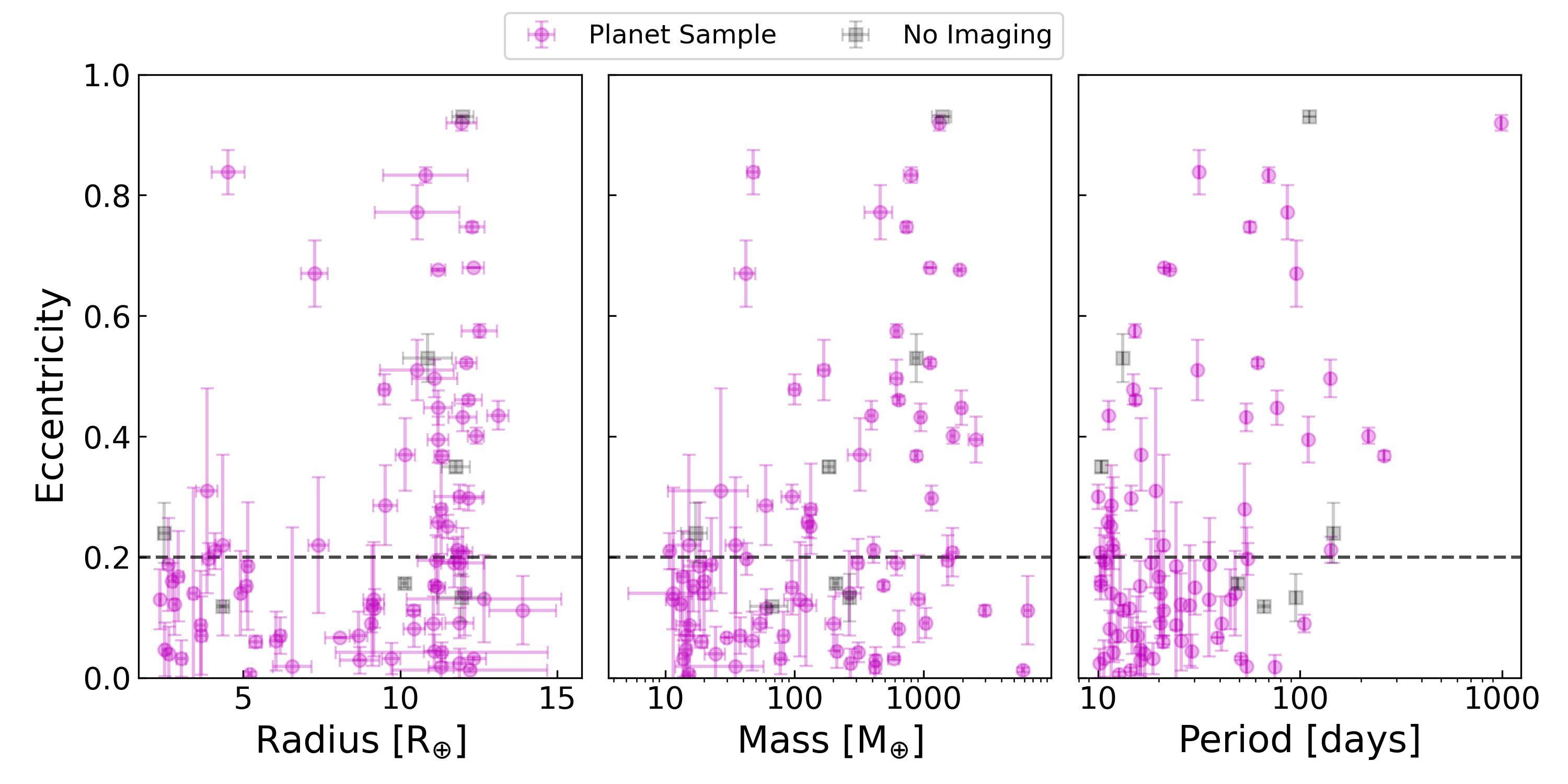} 
  \caption{Radius, Mass and Period vs. Eccentricity (from left to right, respectively) for our target sample. Planets that are part of systems missing high-contrast imaging observations are colored in gray squares.} 
  \label{fig:ecc_plot_missing_imaging}
\end{figure*}

Our sample is comprised of 86 target stars hosting a total of 92 exoplanets, 41\% of which are eccentric (using $e = 0.2$ as the cut off between higher and lower eccentricity systems).
Fig. \ref{fig:ecc_plot_missing_imaging} shows the radius, mass and period vs. eccentricity for the full target sample.
The Radius-Eccentricity plot in the figure shows a separation of our planet sample into two populations around $\sim 6~R_{\oplus}$, while the Period-Eccentricity plot shows a lack of planets at lower eccentricities ($e \leq 0.2$) past $P \sim 150$ days.
Since part of our study explores the eccentricity dependency on the presence of stellar companions (see Section \ref{subsec:stellarcomp}), the availability of high resolution imaging observations plays a significant role in those findings.
Only 7 out of the 86 host stars in our sample are lacking imaging data ($\sim$8\%), which are highlighted in gray in the figure.
We are currently in the process of acquiring high resolution imaging (speckle or adaptive-optics observations) for these targets over the next couple of observing semesters for the purpose of completeness.

Appendix \ref{sec:nexaparams} includes tables of NEXA-derived parameters of our target list that were used in our study.
Table \ref{table:nexaplanetparams} summarizes the planet parameters extracted from NEXA for our sample, and Table \ref{table:nexastarparams} summarizes the stellar parameters of the host stars in our sample.
The reported error bars in the tables correspond to the mean uncertainty of the upper and lower error bounds of the parameters.

For targets missing ages in our sample, we manually checked the publication source to see whether they were available in the literature but were somehow missed by NEXA.
We found a published age of 4 $\pm$ 1 Gyr in the literature for TOI-4582 (TIC 219854519), which we manually added to our table.
Some targets $-$ HAT-P-17 (TIC 266593143), WASP-117 (TIC 166739520), HD 17156 (TIC 302773669), HAT-P-15 (TIC 353459965), Kepler-413 (TIC 298969838), and HD 80606 (TIC 457134360) $-$ were missing stellar metallicity ([Fe/H]) uncertainties in the queried NEXA table, but were found to be available online on the NEXA website and were somehow not getting picked up by the query.
As such, we manually added the missing information to our table.
Kepler-413 (TIC 298969838) was also missing the stellar effective temperature ($T_{\text{eff,$\star$}}$) uncertainty from the queried table despite being available online on NEXA, so we manually updated our table to include it accordingly.

All planet parameters used in this study were taken from our queried NEXA sample described in this section, unless otherwise mentioned.
The NEXA stellar parameters were only used as priors for our homogeneous characterization of the host stars in our sample (see Section \ref{sec:jaxstar}).

\section{Stellar Characterization}\label{sec:jaxstar}

We homogeneously characterize the general stellar properties of our 86 host stars using \texttt{jaxstar}\footnote{\protect\url{https://github.com/kemasuda/jaxstar}}, a python module that provides fast isochrone fitting using HMC-NUTS. 
The method of isochrone fitting was validated using injection-and-recovery tests as well as tests using Kepler seismic stars, which have precise and accurate parameter constraints from asteroseismology \citep[see][for more details]{masuda2022jaxstar}.

To run \texttt{jaxstar}, we used the queried NEXA values and their corresponding errors (see Table \ref{table:nexastarparams}) for the stellar effective temperature $T_{\text{eff,$\star$}}$, metallicity [Fe/H], and $K$ magnitude as priors for the fit, in addition to the precise parallax measurements of our targets from the \textit{Gaia} DR3 catalog.
One system, Kepler-413 (TIC 298969838, highlighted in blue in Fig. \ref{fig:feh_nexa_vs_jaxstar}), had a reported metallicity value of $-$1.44 $\pm$ 0.3 on NEXA, which is unrealistic for an exoplanet host star and is probably a result of a technical error.
As such, we instead use the stellar metallicity [M/H] and $T_{\text{eff,$\star$}}$ from the \textit{Gaia} DR3 catalog \citep{gaia2023dr3} as priors for this target.
The reported values for [M/H] and $T_{\text{eff,$\star$}}$ from \textit{Gaia} DR3 for this source are $-0.39 \pm 0.03$ and 4875$^{+22}_{-18}$ K, respectively. 
For the error bar used in the prior, we took the mean value of the upper and lower error bounds.
Finally, we ran the HMC fit for 20000 warm-up steps and 20000 samples for each host star.

Appendix \ref{sec:jaxstarparams} includes tables and figures related to \texttt{jaxstar}-derived parameters of the host stars in our sample.
Figs. \ref{fig:age_nexa_vs_jaxstar} and \ref{fig:feh_nexa_vs_jaxstar} show the comparison between the NEXA vs. \texttt{jaxstar}-derived stellar ages and metallicities, respectively. 
Fig. \ref{fig:jaxstar_mass_hist} shows the mass distribution of our sample of host stars.
Our homogenously-derived stellar parameters from \texttt{jaxstar} are available in Table \ref{table:jaxstarparams}. 
Four systems $-$ TOI-1278 (TIC 163539739), K2-10 (TIC 363573185), TOI-1231 (TIC 447061717), and Kepler-413 (TIC 298969838) $-$ do not have any reported stellar ages in NEXA or in the literature, and we present their newly-derived \texttt{jaxstar} ages in the table as well.
We achieve a better mean error precision for the stellar ages by $\sim$4\%, with the \texttt{jaxstar} mean error being 1.76 Gyr compared to the NEXA mean error of 1.83 Gyr.
Similarly, we achieve a better mean error precision for the stellar metallicities [Fe/H] by $\sim$11\%, with the \texttt{jaxstar} mean error being 0.056 compared to the NEXA mean error of 0.063.

In our analyses for this study, we use our homogeneous \texttt{jaxstar}-derived stellar parameters wherever relevant or necessary.

\section{Internal Planet Composition}\label{sec:internal}

\begin{figure}
          \centering
          \includegraphics[width=\linewidth]
          {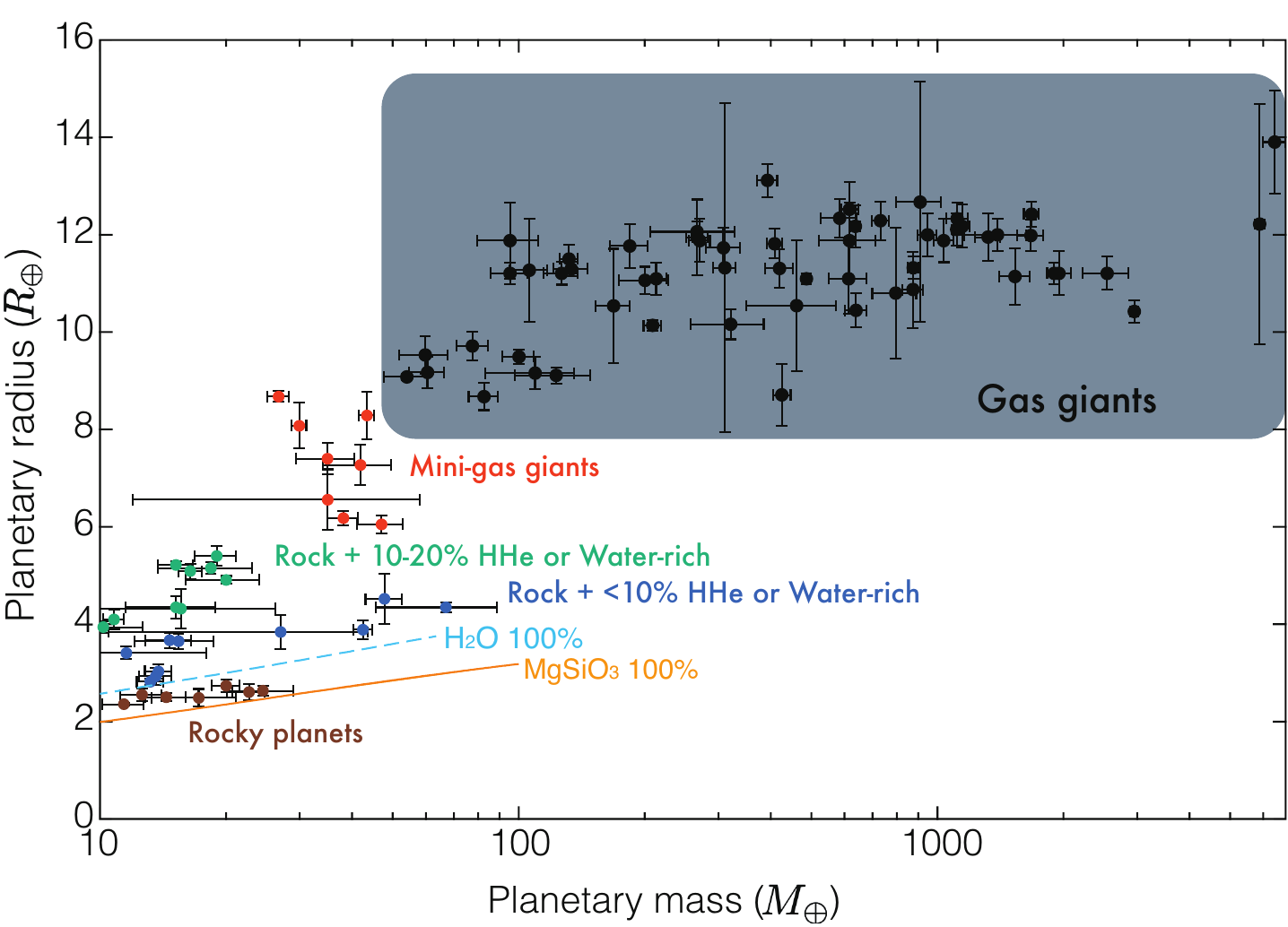} 
          \caption{Mass-radius diagram of our planet sample. Theoretical interior models of pure water planets and pure rocky planets are shown as dashed lines and solid lines, respectively.}
          \label{fig:mr}
\end{figure}

Considering three types of planetary materials (i.e., rock, water ice, and H/He gas), we simulated the interior structures of the planets in our sample using the planet mass, radius and equilibrium temperature.
We used the planetary equilibrium temperature assuming zero Bond albedos.
The interior structure of the planet is integrated using four equations of states (EoSs): the Birch-Murnaghan EoS and the Thomas-Fermi Dirac EoS at  $P > 1.35 \times 10^4$\,GPa of MgSiO$_3$ for rock \citep{seager2007massradius}, AQUA EoS for water ice \citep{haldemann2020aqua}, and H/He \citep{chabrier2021eqofstate}.
We found that our sample of planets can be categorized by the following 5 groups based on their internal composition (see Figure \ref{fig:mr}):
\begin{enumerate}
    \item Gas giant planets ($R_{\text{p}}$ > 8 $R_{\oplus}$ and $M_{\text{p}}$ > 50-60 $M_{\oplus}$)
    \item Mini-gas giant planets ($R_{\text{p}}$ > 6 $R_{\oplus}$ and $M_{\text{p}}$ $\sim$30-50 $M_{\oplus}$)
    \item Rocky planets with 10-20\% H/He envelope or Water-rich planets ($R_{\text{p}}$ > 4 $R_{\oplus}$ and $M_{\text{p}}$ $\sim$10-20 $M_{\oplus}$)
    \item Rocky planets with < 10\% H/He envelope or Water-rich planets ($R_{\text{p}}$ $\sim$3-4 $R_{\oplus}$)
    \item Rocky planets with < a few \% H/He ($R_{\text{p}}$ < 3 $R_{\oplus}$)
\end{enumerate}

Note that the mass-radius relation of small planets allows two solutions for their internal composition: water-rich planets and rocky planets surrounded by H/He envelopes.
We use these five groups to help us parametrize the modified tidal quality factor $Q_{\text{p}}$ in Section \ref{subsubsec:tidalquality} when calculating the tidal dissipation timescale.

\section{Eccentricity Distribution}\label{sec:eccdist}

\begin{figure}
  \centering
  \includegraphics[width=\linewidth]
  {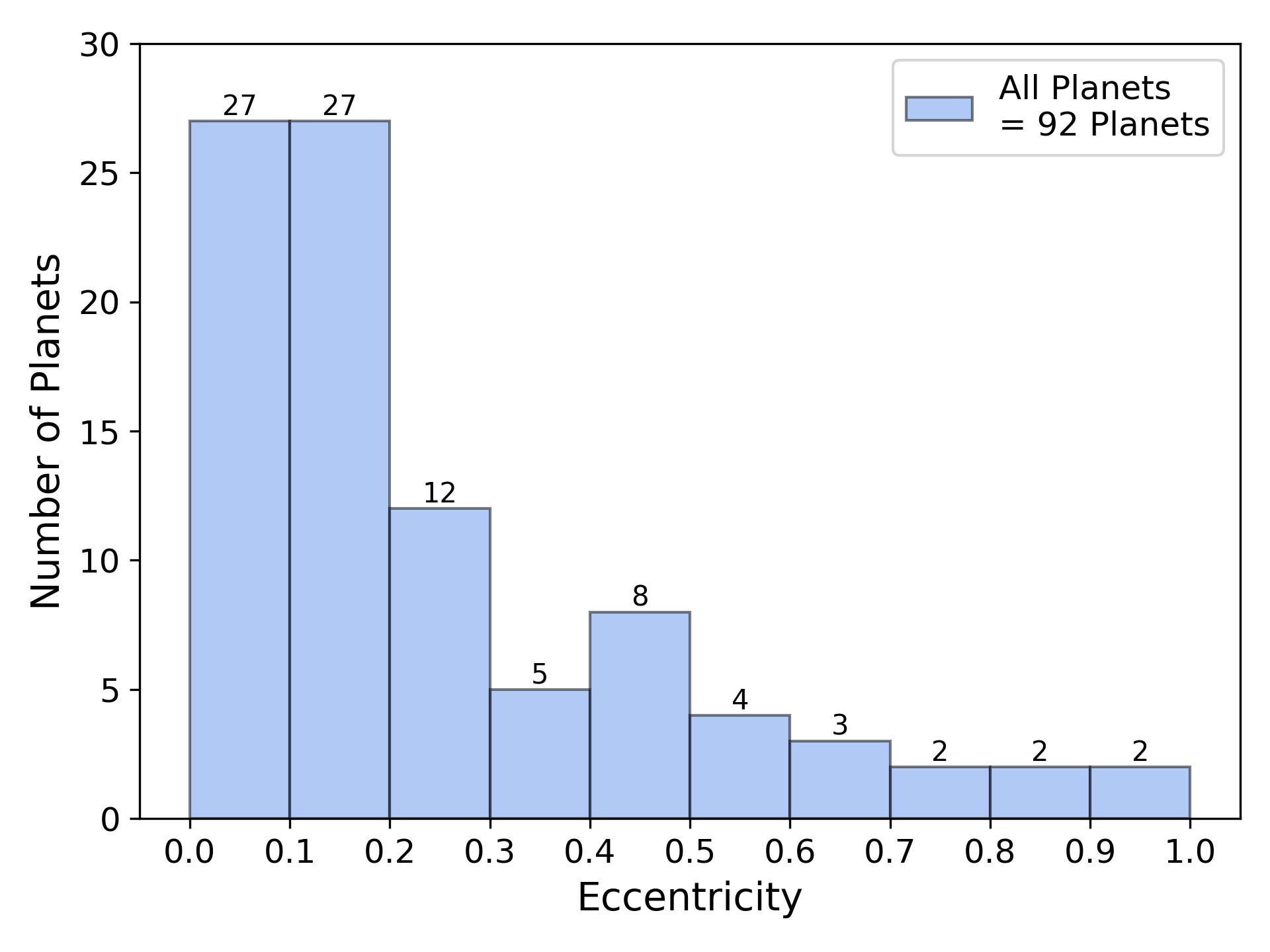} 
  \caption{Eccentricity distribution of all 92 planets in our sample.} 
  \label{fig:ecc_hist_all}
\end{figure}

In this paper, we explore the eccentricity distribution of TLGs and investigate their correlation with different planetary and stellar factors, which could provide useful insights into the eccentricity origin of our planet sample.
We use the eccentricity measurements queried from NEXA, as described in Section \ref{sec:sample}.
Fig. \ref{fig:ecc_hist_all} shows the eccentricity distribution of all 92 planets in our target list.
In the following sections, we divide our sample based on different criteria to see whether the inherent eccentricity distribution is dependent on such factors.
More specifically, we test the eccentricity correlation of TLGs with: stellar age, stellar metallicity, stellar companion, planet radius, planet multiplicity, planet equilibrium temperature and planet tidal dissipation timescale (in relation to stellar age).

To test the significance of our eccentricity distributions across different parameter spaces, we perform the Kolmogorov-Smirnov (K-S) test using the built-in \texttt{kstest} function from the \texttt{scipy} python module \citep{virtanen2020scipy}. 
This enables us to compare sub-samples of the eccentricity distribution according to different cuts and test the null hypothesis for whether they are distributed according to the standard normal. 
We discuss the selection criteria used to split our sample and perform the K-S tests in each subsequent section separately, along with the implications of our results.
Additionally, we use the Spearman correlation coefficient (also known as the Spearman $\rho$ test) to verify the existence of a correlation between two continuous parameters that are not split by categories.
We use the built-in \texttt{spearmanr} function from the \texttt{scipy} python module \citep{virtanen2020scipy}.
Since this is a non-parametric test, it does not make assumptions on the specific form of the correlation, and allows us to analytically calculate the probability that a given value of the Spearman coefficient comes by chance when there is no correlation. 
While the $p$-value calculation in this test does not make strong assumptions about the distributions underlying the samples, it is only accurate for very large samples (> 500 observations).
For smaller samples, it is more appropriate to perform a permutation test, where one can produce an exact null distribution by calculating the statistic under each possible pairing of elements between the two continuous parameters.
Since our sample is small and consists of 92 planets, we perform a permutation test when calculating the $p$-value for the Spearman $\rho$ test.

For continuous parameters, the K-S test requires us to fix a boundary value for one of the two variables to construct the two distributions to be compared. 
Since this choice is not unique in most cases, we chose to perform Spearman $\rho$ tests when both variables are continuously varying. 
On the other hand, the K-S method is very useful and appropriate when there is a natural way to define different distributions to be compared based on categories (e.g. stars with detected distant companions and star without detected companions).
Using a confidence level of 95\%, we consider a parameter to be statistically significant when tested for correlations with eccentricity if both the K-S test and the Spearman $\rho$ test have $p$-values less than 0.05 (when the parameter is not continuous, we only consider the K-S test).

\section{Results and Discussion}\label{sec:results_discussion}

In this work, we focused on a sample of 92 transiting, long-period giant planets as part of an eccentricity distribution study for this planet population in order to understand their eccentricity origin.
Our main focus in the present paper is to look for trends and possible correlations in the eccentricity distribution, and the detailed comparisons with theoretical models/expectations will be given in future works.
Past studies have performed population level analyses for the eccentricity distributions of small planets \citep{kane2012eccdistkepler, xie2016ecckepler, vaneylen2019eccdist} and close-in giant planets \citep{knutson2014friendshj, bonomo2017gaps}, but long-period giant planets remain widely unexplored.
In particular, previous studies of giant planets focused on hot jupiters, which are substantially different to TLGs and have different characteristics and formation pathways. 
As such, direct comparisons are difficult to make in our specific case since the eccentricity distributions of TLGs have not been well-explored before, but we highlight any relevant comparisons that could be made when possible.

In the following subsections, we present our findings of the eccentricity distributions in relation to stellar age, stellar metallicity, stellar companion, planet radius, planet multiplicity, planet equilibrium temperature and tidal dissipation.
We then discuss the implications of our findings for each correlation (or lack of) with the eccentricity distribution of TLGs.
The results of the parameters tested for statistical correlations against eccentricity are presented in Table \ref{table:statstests}.

\renewcommand*{\arraystretch}{1.3}
\begin{table*}
\caption{Results of the statistical tests used to determine which parameters are correlated with eccentricity. The thresholds listed are only used for the K-S tests, since the sample had to be split based on a cut-off threshold in order to compare the two distributions. Parameters with no listed thresholds were divided by category (e.g. companion vs. no companion, single vs. multi). They were also not tested under the Spearman $\rho$ Test since they are not continuous parameters. $p$-values that were found to be < 0.05 are highlighted in bold.}
\label{table:statstests}
    \centering
    \begin{tabular}{lccc}
    \hline
    Parameter Name                                     & Threshold & $p$-value (K-S Test) & $p$-value (Spearman $\rho$ Test) \\
    \hline
    Stellar Age [Gyr]                                  & 5         & 0.930                & 0.442                            \\
    Stellar Metallicity [Fe/H]                         & 0.1       & \textbf{0.022}       & \textbf{0.040}                   \\
    Stellar Companion                                  & --        & 0.956                & --                               \\
    Stellar Effective Temperature $T_{\text{eff}}$ [K] & 5629      & 0.494                & 0.250                            \\
    Planet Period [days]                               & 21        & 0.211                & 0.070                            \\
    Planet Radius $R_{\text{p}}$ [$R_{\oplus}$]        & 6         & \textbf{0.030}       & \textbf{0.005}                   \\
    Planet Multiplicity $N$                            & --        & \textbf{0.048}       & --                               \\
    Planet Equilibrium Temperature $T_{\text{eq}}$ [K] & 500       & 0.973                & 0.676                            \\
    Timescale Ratio $\tau$ ($Q_{\text{p}}$ = 10$^{2}$) & 0.9       & \textbf{0.035}       & 0.089                            \\
    Timescale Ratio $\tau$ ($Q_{\text{p}}$ = 10$^{3}$) & 0.9       & 0.279                & 0.089                            \\
    Timescale Ratio $\tau$ ($Q_{\text{p}}$ = 10$^{4}$) & 0.9       & 0.669                & 0.089                            \\
    Timescale Ratio $\tau$ ($Q_{\text{p}}$ = 10$^{5}$) & 0.9       & 0.316                & 0.089                            \\
    Timescale Ratio $\tau$ (Varying $Q_{\text{p}}$)    & 0.9       & 0.310                & \textbf{0.018}                   \\
    \hline
    \end{tabular}
\end{table*}

    \subsection{Stellar Age}\label{subsec:stellarage}
    
\begin{figure}
      \centering
      \includegraphics[width=\linewidth]
      {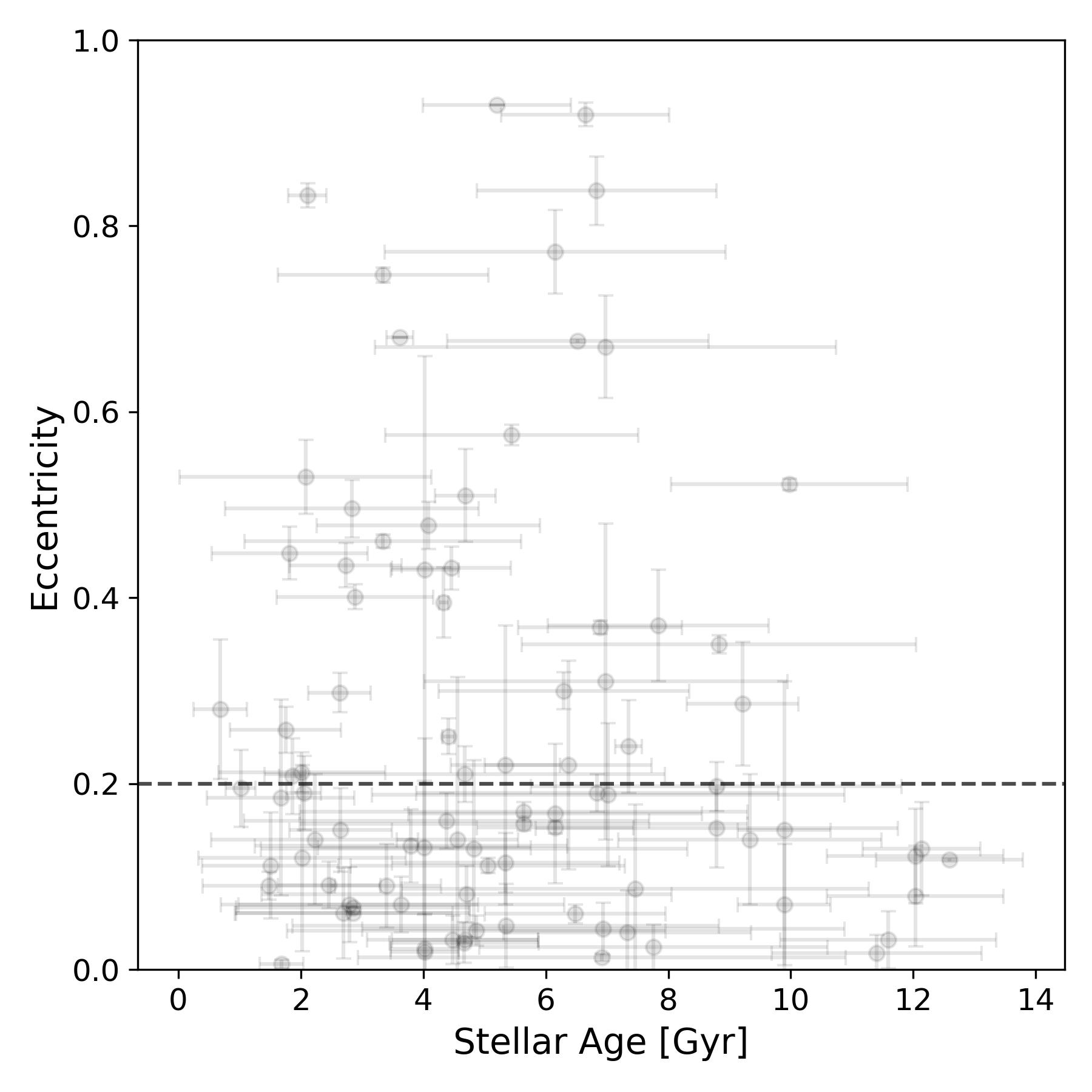} 
      \caption{Stellar Age vs. Eccentricity for all 92 planets in our sample.} 
      \label{fig:jaxstar_age_vs_e}
    \end{figure}

    We explore the stellar age correlation to the planet eccentricity for our sample of TLGs.
    Fig. \ref{fig:jaxstar_age_vs_e} shows the stellar age vs. eccentricity for all 92 planets in our sample.
    There are no visible trends or correlations by eye, and the ages appear to be distributed relatively homogeneously.
    Although, we note a lack of high-$e$ planets around older host-stars (Age > 10 Gyr), which could suggest that these planetary systems have settled down.
    
\begin{figure}
      \centering
      \includegraphics[width=\linewidth]
      {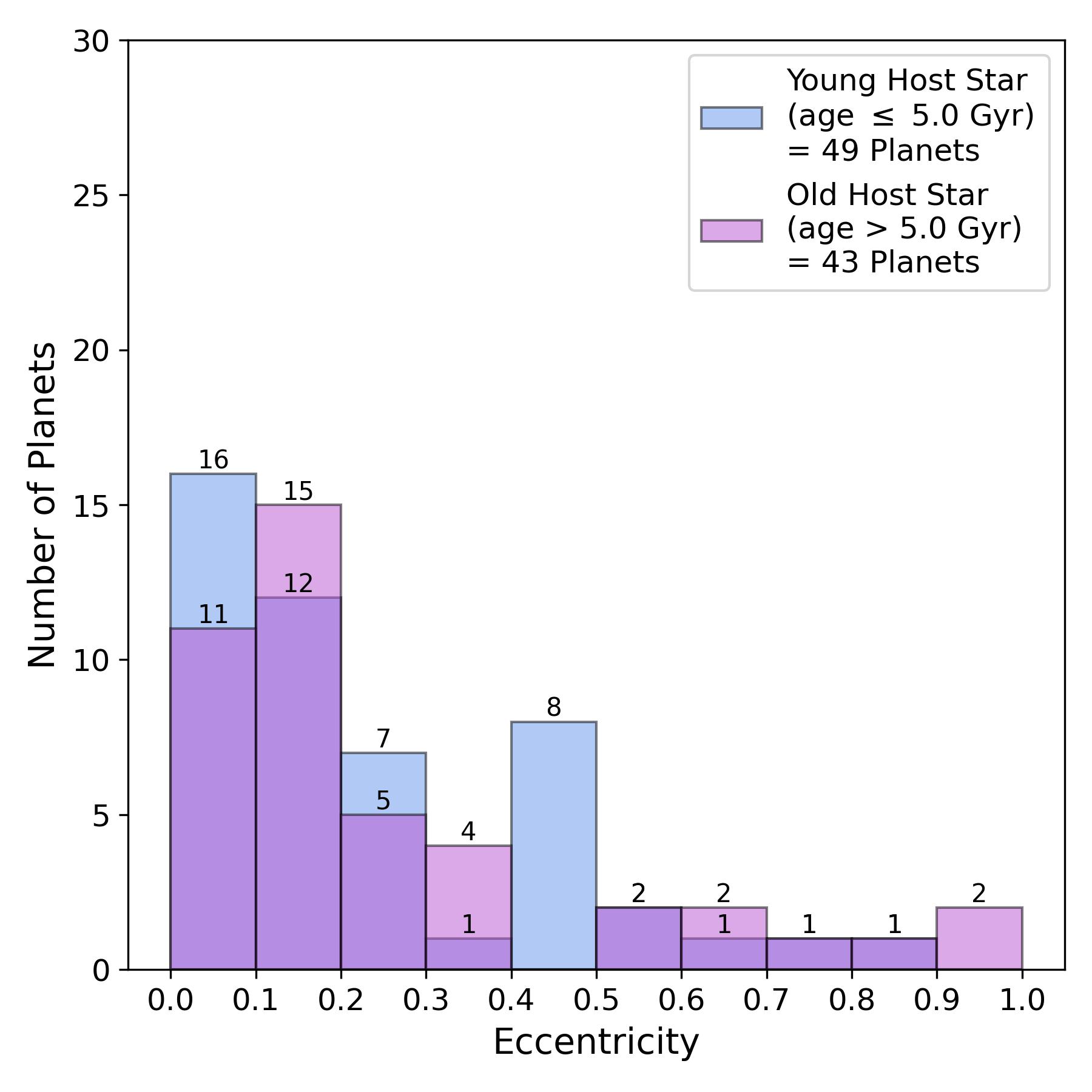} 
      \caption{Eccentricity distribution of planet systems with young (blue) vs. old (magenta) host stars in our sample, using 5 Gyr as the cut-off threshold between the two sub-samples.} 
      \label{fig:ecc_hist_young_vs_old}
    \end{figure}

    To check for statistical correlations using the K-S test, we split our sample based on whether planets belong to young or old host-stars, using 5 Gyr as the cut-off threshold between the two sub-samples.
    This threshold was chosen based on the median age of our sample.
    Fig. \ref{fig:ecc_hist_young_vs_old} shows the eccentricity distribution of planet systems with young (blue) vs. old (magenta) host stars in our sample.
    Both results from the K-S test and the Spearman $\rho$ test have $p$-values > 0.05, indicating that there is no statistical correlation between age and eccentricity. 
    
    The lack of correlation between host-star age and planet eccentricity for our sample TLGs is not surprising. 
    Previous studies of eccentricity distributions do not report any correlation with stellar age, even across planet groups beyond TLGs \citep{udry2007exostats, bowler2020eccdist}.
    Interestingly, we only find young host-star planets in the 0.4-0.5 eccentricity bin in Fig. \ref{fig:ecc_hist_young_vs_old}, but the lack of old host-star systems in this regime is likely a result of the small sample size beyond $e > 0.2$, whereas the sample size for young host-star planets only starts to decrease significantly beyond $e > 0.4$.
    Thus, we cannot determine whether this observation is to be attributed to a sampling issue or has a physical explanation related to the host-star age.
    
    \cite{swastik2023age} determined the age distribution of exoplanet host-stars and found that stars hosting giant planets tend to be younger than stars hosting small planets, but the study did not explore whether there was any correlation with the planet eccentricity.
    We also note that stellar ages are generally difficult to estimate and usually have very large error bars. 
    While we managed to derive homogeneous measurements of stellar ages and achieved better error precisions than the ages reported on NEXA, there are still issues with model degeneracies when performing isochrone fitting.
    Other methods of determining ages (such as astroseismology) have been shown to provide more accurate measurements \citep{silva2015ageastroseis, aerts2021astroseis}, but astroseismic data are limited, making it difficult to utilize in statistical studies without compromising on the sample size.   

    In younger systems, planets can form with a diverse range of eccentricities, and can start off with very high eccentricities due to the chaotic interactions and processes during the early stages of formation. 
    As the system stabilizes with time and age, such planets (especially ones that formed or migrated close to the host star) will have undergone tidal dissipation and began to circularize, causing eccentricities to be dampened to lower values over time \citep{villaver2014hotjup}.
    So, as the system reaches older ages, we would expect to see fewer planets with very high eccentricities.
    In our case, given the period cut we placed at 10 days, the planets in our sample are too far from the host star for tidal dissipation effects to cause any significant eccentricity damping.
    As a sanity check, we tested whether there was any period correlation with eccentricity given the period cut we imposed.
    Both results from the K-S test and the Spearman $\rho$ test have $p$-values > 0.05, indicating that there is no statistical correlation between period and eccentricity.
    As such, we do not expect age to play a significant role on the observed eccentricity in our planet sample, which is also confirmed by our findings.
    A more detailed discussion on the effects of tidal dissipation for our sample can be found in Section \ref{subsec:tidaldissipation}.

    \subsection{Stellar Companion}\label{subsec:stellarcomp}
    
    We explore the correlation between the presence of stellar companions to the planet eccentricity for our sample of TLGs. 
    
    To find systems with a nearby stellar companion, we manually searched the Exoplanet Follow-up Observing Program (ExoFOP) website\footnote{\protect\url{https://exofop.ipac.caltech.edu}} for each target and checked whether high resolution imaging observations were taken, and if so, whether there was any evidence of detected companions reported. 
    We placed no constraints on the separation of the companion, and we caution that this approach depends on the detection limits of the instrument and telescope used, which is heterogeneous across our sample.
    On ExoFOP, our targets had Speckle, AO or Lucky imaging, and in some cases, they were observed by more than one imaging technique.
    Some of the AO imaging observations used in this study will be published as part of a catalog paper by Dressing et al. (submitted).
    The AstraLux Lucky-imaging data used can be found in \cite{lillobox2024astralux}.
    As highlighted earlier in the paper, only $\sim$8\% of the host-stars in our sample are lacking imaging observations.
    Since this is a relatively small number of systems, and they are also homogeneously distributed across different eccentricity ranges (see Fig. \ref{fig:ecc_plot_missing_imaging}), this lack of imaging data should not significantly affect our findings.
    14 out of the 92 planets in our sample ($\sim$15\%) were found to have a stellar companion detected in the high-resolution imaging data.
    
    To find systems with a wide, co-moving companion, we utilized the \cite{elbadry2021bincatalogue} catalog, which made use of \textit{Gaia} eDR3 \citep{gaia2021edr3} to find spatially resolved binary stars within $\sim$1 kpc of the Sun, with projected separations ranging from a few AU to 1 pc.
    12 out of 92 planets in our sample ($\sim$13\%) belonged to systems containing a wide stellar companion when cross-matched with the \cite{elbadry2021bincatalogue} catalog.

    The presence of nearby stellar companions has been shown to influence the measured planet parameters \citep{furlan2017compeffect} and stellar parameters \citep{furlan2020compeffect}, in particular with respect to radius.
    Discovery papers normally take into account the dilution effects on the measured planet and stellar parameters, assuming that the stellar companion was already found at the time of publication.
    The situation becomes less clear for systems where a nearby stellar companion was only found after the discovery paper was published.
    To see whether this would pose a problem for companion-detected systems in our sample, we first checked the projected physical separation of the companion candidates, and found that all detected stellar companions (whether found by high-resolution imaging or in the El-Badry catalog) were relatively distant from the exoplanet host star, with separations larger than 100 AU.
    Additionally, if the companions have angular separations > 5\arcsec, they are usually identified in the \textit{Gaia} database, and taken into account to estimate the stellar parameters (TIC catalog) as well as the radius ratio ($R_{\text{p}}$/$R_{\star}$) in the transit modeling (TOI catalog). 
    
    Next, for stellar companions with angular separations < 5\arcsec, we checked their magnitude difference $\Delta M$ with respect to the host star to see whether they could pose any contamination issues.
    We found that the majority of the stellar companions detected around our targets are relatively faint ($\Delta M > 5$) and only 3 targets (TOI-2589, TOI-2010 and Kepler-434) had stellar companions that were brighter than $\Delta M = 4$.
    The discovery papers of TOI-2589 \citep{brahm2023toi2589} and TOI-2010 \citep{mann2023toi2010} were aware of the stellar companions and took into account the dilution effect when estimating the stellar and planetary parameters.
    While the discovery paper of Kepler-434 \citep{almenara2015kepler434} did not account for the stellar companion, the system was revisited by \cite{berger2018kepler434revised}, who revised Kepler planet radii using \textit{Gaia} DR2, and reported that only low-contrast companions with separations < 4\arcsec could dilute the measured fluxes enough to exceed their reported 8\% uncertainties.
    Given that the companion of Kepler-434 is bound, and its $\Delta M_{J} = 3$ should correspond to $\Delta M_{K_p} > 4$, it is a relatively high-contrast companion and its dilution effect is almost negligible given the radius uncertainty (as well as uncertainties in stellar parameters).
    
\begin{figure}
      \centering
      \includegraphics[width=\linewidth]
      {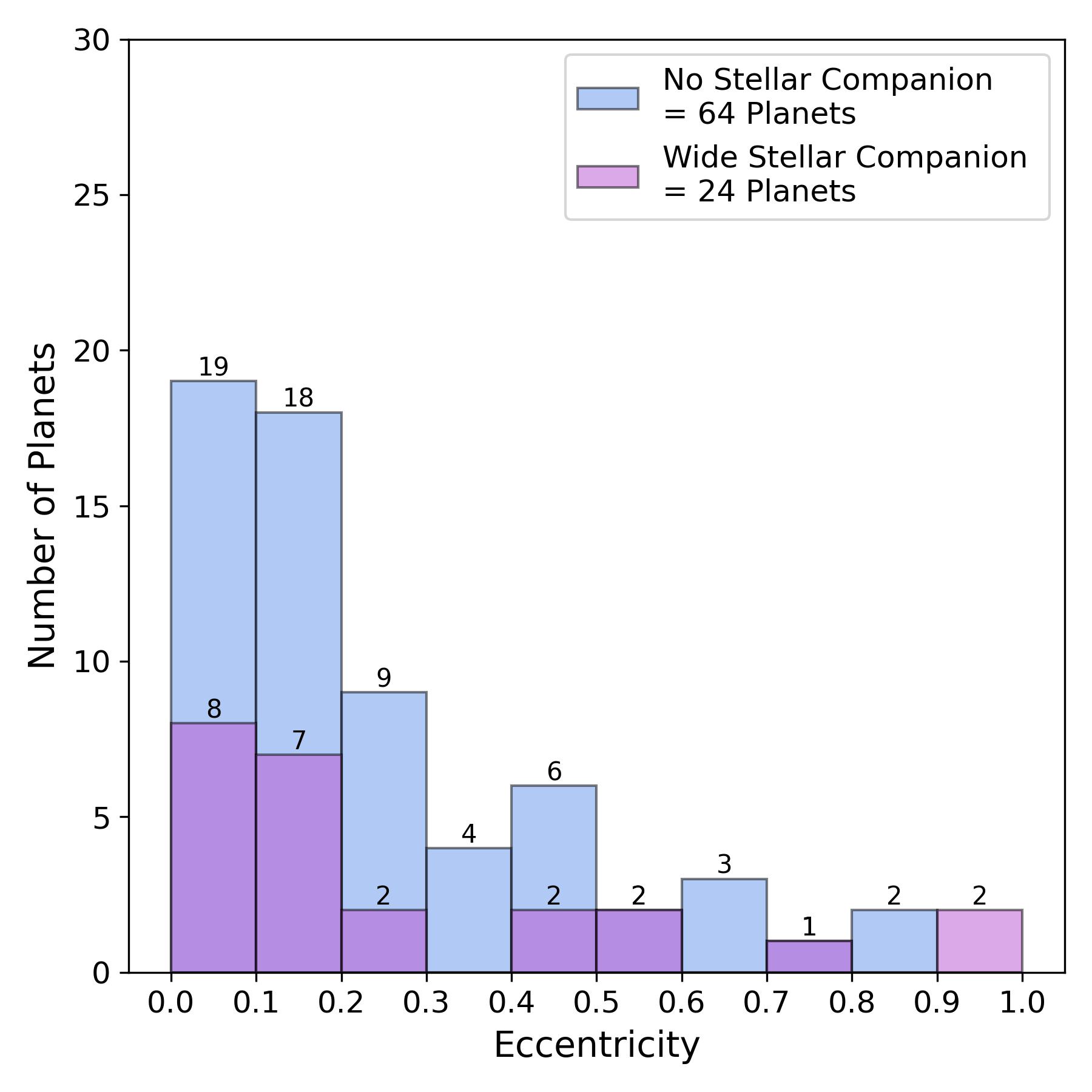} 
      \caption{Eccentricity distribution of planets with no detected stellar companions (blue) vs. planets with wide stellar companions (magenta) in our sample. 4 out of the 92 planets ($\sim$4\%) were excluded from this sample due to lack of imaging data.} 
      \label{fig:ecc_hist_no_comp_vs_closein_OR_wide}
    \end{figure}

    Fig. \ref{fig:ecc_hist_no_comp_vs_closein_OR_wide} shows the eccentricity distribution of planets with no detected stellar companions (blue) vs. wide stellar companions (magenta) in our sample.
    In the no-companion distribution, we excluded planets that had no data available since the presence of companions is uncertain in these cases. 
    If systems lacked imaging observations but were found to have a wide companion, they were not excluded. 
    In summary: systems were only excluded if they lacked high-resolution imaging observations and were also not found in the \cite{elbadry2021bincatalogue} catalog.
    This resulted in a total of 4 out of the 92 planets ($\sim$4\%) being excluded due to lack of data.
    We find a $p$-value > 0.05 from the K-S test, which indicates that there is no statistical correlation between the planet eccentricity and the presence of stellar companions for TLGs.
    This could provide important clues about the eccentricity origin for our sample.
    
    Previous studies of eccentricity distributions of small planets and close-in giant planets similarly found no statistical trend with stellar companions.
    \cite{vaneylen2019eccdist} reported that there was no noticeable difference in the eccentricity distributions of small planets around single stars and those orbiting a star with a close stellar companion.
    \cite{knutson2014friendshj} also found no eccentricity dependence of close-in giant planets on the presence of stellar companions, as well as no observed difference in the frequency of companions for planets with well-aligned circular orbits and misaligned eccentric orbits.
    The lack of correlation between the planet eccentricity and the presence of wide stellar companions in our sample indicates that these warm, long-period giant planets may have experienced secular planet-planet scattering, and that the Kozai-Lidov mechanism does not play a critical role in the eccentricity distribution of TLGs.
 
    If more than one giant planet is initially formed in the system, then planet-planet interaction could eject one giant planet out of the system while leaving behind an eccentric giant planet within the system \citep{lin1997originecc}.
    Additionally, \cite{dawson2013planetscat} showed strong evidence that gas giants with higher eccentricities (in particular ones orbiting higher metallicity host-stars) are driven by the presence of another giant planet.
    This is supported by previous works that investigated whether the highly eccentric exoplanet population can be produced entirely by scattering \citep{ford2008planetscat, carrera2019planet}. 
    \cite{carrera2019planet} showed that the eccentricity distribution for giant planets with $e>0.3$ was found to be consistent with the planet-planet scattering scenario, and they highlight how the Kozai-Lidov mechanism is not necessarily the default source of the eccentricity origin for planets discovered with very high eccentricities.

    As a sanity check, we roughly estimated the Kozai-Lidov oscillation timescale for systems with confirmed stellar companions \citep[see][]{holman1997chaotic, shevchenko2020kozailidov}, and found that for a majority of the targets (>60\%), the timescale is longer than the inferred age of the system, which may be partly responsible for the absence of a correlation between eccentricity and the presence of a stellar companion.
    It is worth noting though that Kozai-Lidov is predicted to produce planets with high eccentricities and low mutual inclinations, or low eccentricities and high mutual inclinations \citep{hatzes2016architecture}.
    In such cases, it could still be possible for the TLGs in our sample to have acquired their eccentricities via the Kozai-Lidov effect.  
    \cite{bowler2020eccdist} used hierarchical Bayesian modeling to test for population-level trends in the stellar companion eccentricity distributions of 27 long-period giant planets and brown dwarfs, where they found significant differences when looking at companion mass and mass ratio.
    They reported that the stellar companions of giant planets have a preference for low eccentricities ($e \sim 0.05-0.25$), which provides evidence for in situ formation on largely undisturbed orbits within massive extended disks.
    If the Kozai-Lidov mechanism is the source of the eccentricity origin for such planets in our sample, we might expect these companions to have lower eccentricities.
    Further follow-up studies of the mutual inclinations of planets in our sample could help shed light on the prominence of Kozai-Lidov and whether it does indeed play any role, but this is beyond the scope of this paper.

    \subsection{Planet Radius}\label{subsec:radius}
    
\begin{figure}
          \centering
          \includegraphics[width=\linewidth]
          {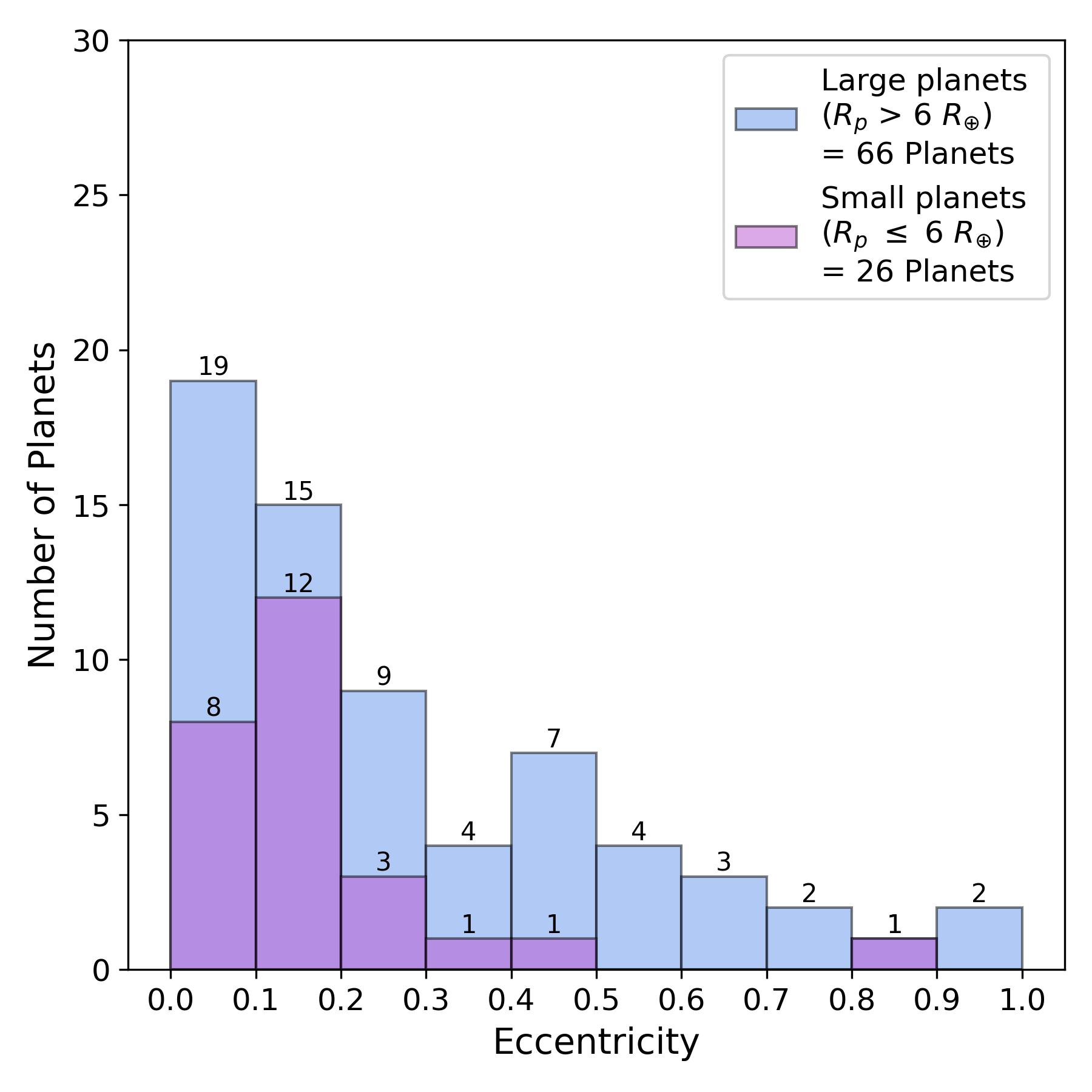} 
          \caption{Eccentricity distribution of large (blue) vs. small (magenta) planets, using a cut-off threshold of $R_{\text{p}}$ = 6 $R_{\oplus}$.} 
          \label{fig:ecc_hist_small_vs_large}
        \end{figure}

    We explore the dependence of the eccentricity distribution of TLGs on the planet radius $R_{\text{p}}$.
    There is an interesting gap in the radius distribution at $\sim 6~R_{\oplus}$ (see Fig. \ref{fig:ecc_plot_missing_imaging}), indicating the possibility of two separate populations within our sample.
    This is further supported by the results of our internal structure modeling (see Section \ref{sec:internal}), where planets with $R_{\text{p}}$ > 6 $R_{\oplus}$ were found to be gas giants and compositionally different from the smaller planet counteparts below 6 $R_{\oplus}$.
    To test the significance of this using the K-S test, we split our sample based on the planet radius, using a cut-off threshold of $R_{\text{p}}$ = 6 $R_{\oplus}$. 
    Fig. \ref{fig:ecc_hist_small_vs_large} shows the eccentricity distribution of large (blue) vs. small (magenta) planets in our sample.
    Both results from the K-S test and the Spearman $\rho$ test have $p$-values < 0.05, indicating that there is a statistical correlation between $R_{\text{p}}$ and eccentricity for TLGs. 
    We can see that the majority of small planets have lower eccentricities, and planets with higher eccentricities tend to be larger.
    
    We found that planet radius plays a significant role in the eccentricity distribution of TLGs, where small vs. large planet systems produced statistically different distributions (see Fig. \ref{fig:ecc_hist_small_vs_large}), indicating that these two populations are separate and could have different sources for their eccentricity origin. 
    Eccentricity distributions of small planets similarly found a radius correlation.
    \cite{kane2012eccdistkepler} reported a radius-dependence for \textit{Kepler} candidates, where smaller planets were found to have lower eccentricities.
    It is important to note that having a mixture of planet groups in our sample (see our internal composition modeling in Section \ref{sec:internal}) introduces issues such as the planet composition degeneracy for sub-Neptunes, where it is very difficult to differentiate between a rocky core with a gaseous envelope vs. a planet with a significant water mass fraction in its atmosphere (water worlds).
    The formation pathways and eccentricity evolution also vary across different planet groups and will not be the same for all planets in our sample.
    This highlights the benefit of placing more strict radius cuts for future studies in order to separate planet groups with different internal compositions.

    \subsection{Planet Multiplicity}\label{subsec:multiplicity}
    
\begin{figure}
      \centering
      \includegraphics[width=\linewidth]
      {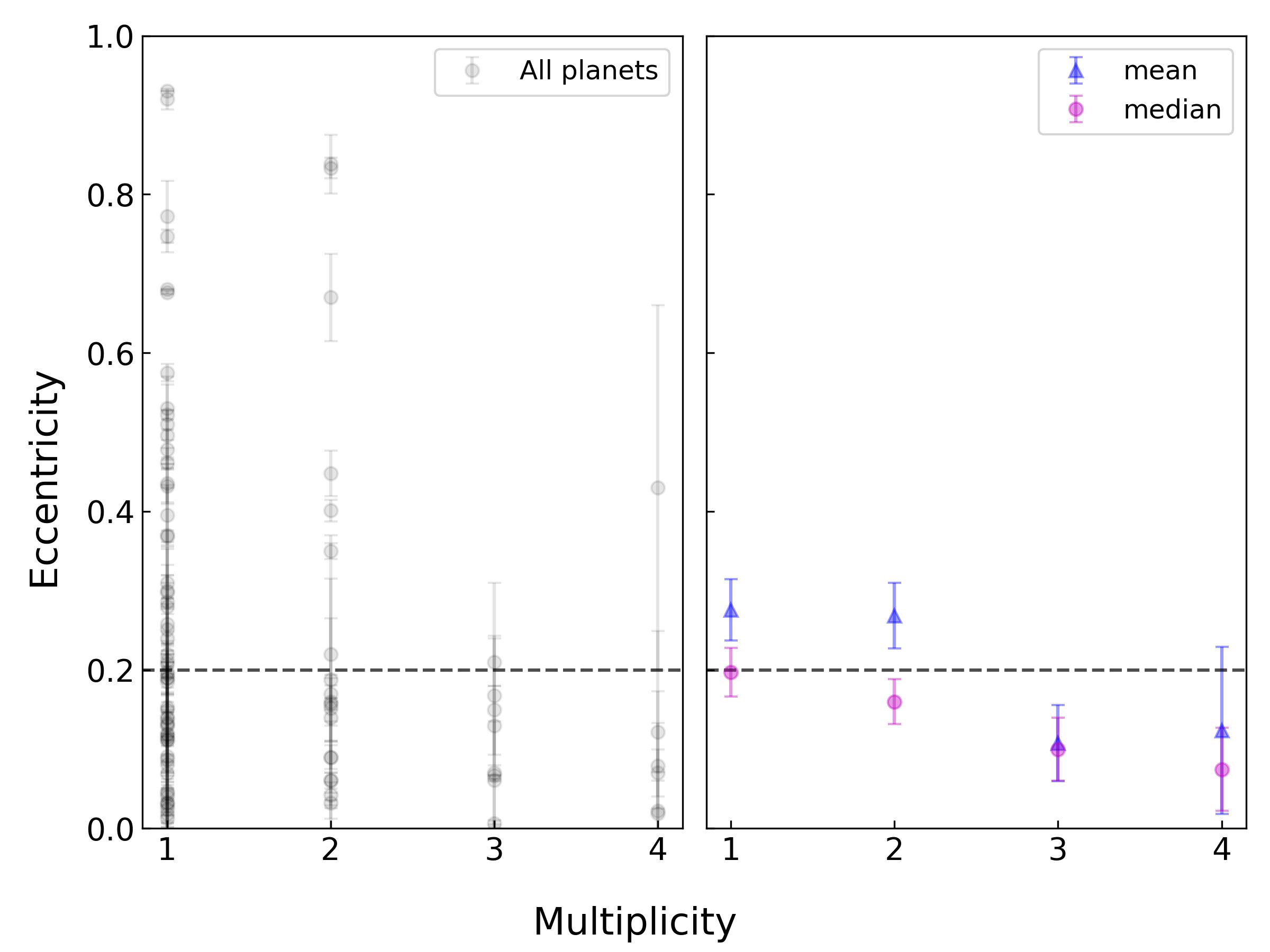} 
      \caption{Planet Multiplicity vs. Eccentricity for targets in our sample. The left panel shows all the planets in our sample, while the right panel shows the weighted mean and median eccentricity values for each multiplicity bin.} 
      \label{fig:n_planets_full_stats_vs_e}
    \end{figure}

    We explore the relation of planet multiplicity $N$ with eccentricity $e$ in our sample of TLGs.
    Fig. \ref{fig:n_planets_full_stats_vs_e} shows the planet multiplicity vs. eccentricity for targets in our sample. 
    The largest eccentricities appear to be dominated by single-planet ($N=1$) and 2-planet ($N=2$) systems, after which the planet eccentricities quickly fall closer to $e \leq 0.2$ for $N>2$.
    
    Previous empirical studies have shown an anti-correlation trend between $N$ and $e$, usually described by a power law in the form of $e(N) = \alpha \times N^{\beta}$ \citep{limbach2015multiplicity, zinzi2017multiplicity, bachmoller2021multiplicity}.
    It is worth noting the different selection criteria adopted by these studies.
    \cite{limbach2015multiplicity} focused on cataloged radial velocity (RV) systems for their sample.
    \cite{zinzi2017multiplicity} limited their sample to planets that were around stars with effective temperatures between 2600 K and 7920 K, in systems with at least two planets, and were discovered with either the RV or transit methods.
    \cite{bachmoller2021multiplicity} did a larger-scale study and included all confirmed planets listed on NEXA, regardless of detection method. 
    The only requirement they placed was for the eccentricity to have error measurements listed on NEXA.
    As part of their study, they also split their sample into subsets to explore possible correlations (e.g. with planet types and/or detection methods), and found that all sub-samples consistently followed the same basic trend. 
    Our choice to only focus on TLG planets makes our sample distinct in comparison to previous works, all of whom placed no criteria on planet types.
    
\begin{figure}
      \centering
      \includegraphics[width=\linewidth]
      {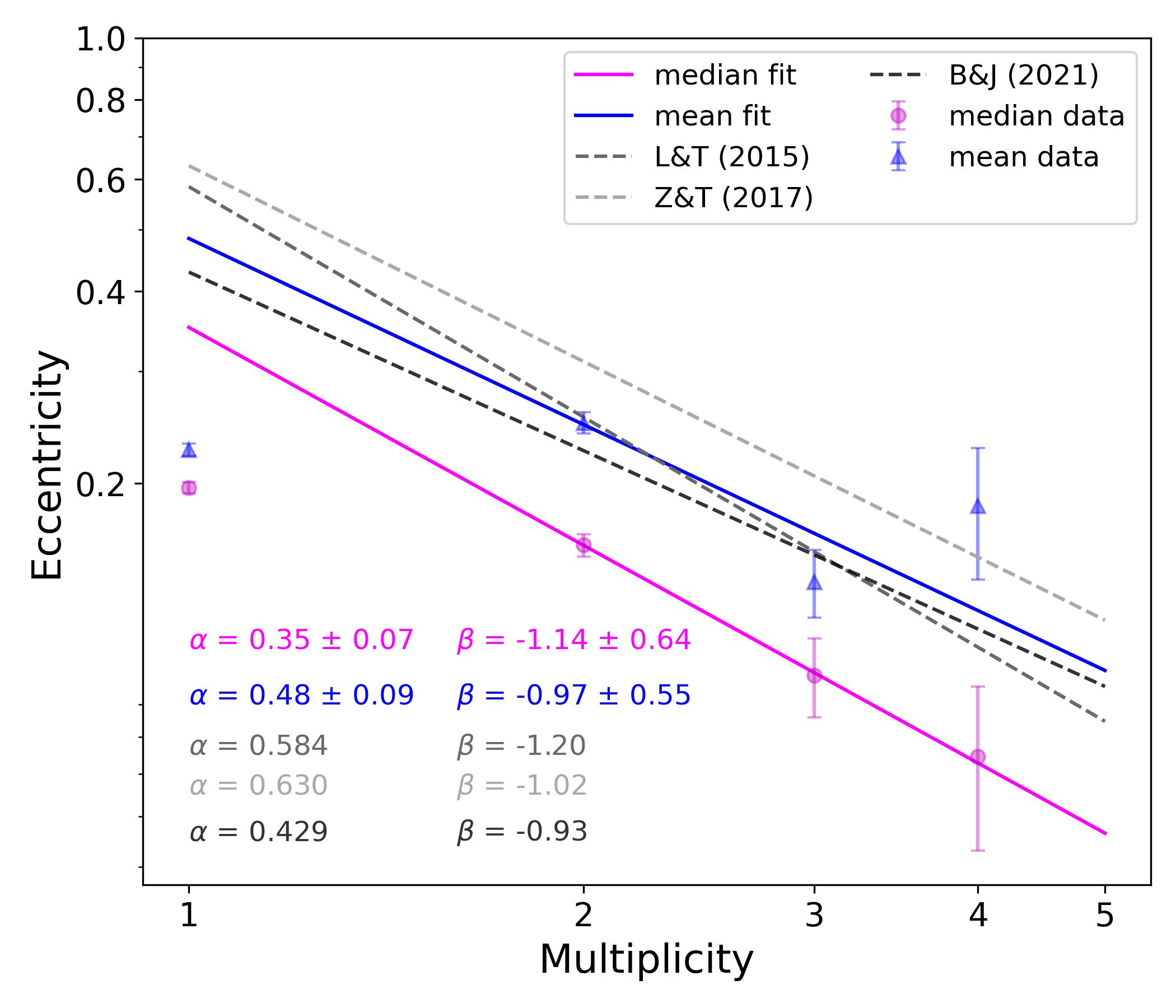} 
      \caption{Planet Multiplicity vs. Eccentricity for targets in our sample. The solid lines are the best-fit power-law models to the weighted mean (blue) and median (magenta) eccentricities of our sample (for $N$ > 1). For comparison, we also over-plot the power law trends from \protect\cite{limbach2015multiplicity} (gray), \protect\cite{zinzi2017multiplicity} (light-gray), and \protect\cite{bachmoller2021multiplicity} (black) using dashed lines.} 
      \label{fig:n_planets_above_1_vs_e_powerlaw_fit}
    \end{figure}

    Fig. \ref{fig:n_planets_above_1_vs_e_powerlaw_fit} shows the planet multiplicity $N$ vs. eccentricity $e$ for targets in our sample, along with our best-fit power law models to the weighted mean and the median eccentricities.
    We also over-plot the power law trends from previous studies to compare with our best-fit models.
    Our best-fit power law to the weighted mean and the median eccentricities is found to be $e(N) = 0.35 \times N^{-1.14}$ and $e(N) = 0.48 \times N^{-0.97}$, respectively. 
    
\begin{figure}
      \centering
      \includegraphics[width=\linewidth]
      {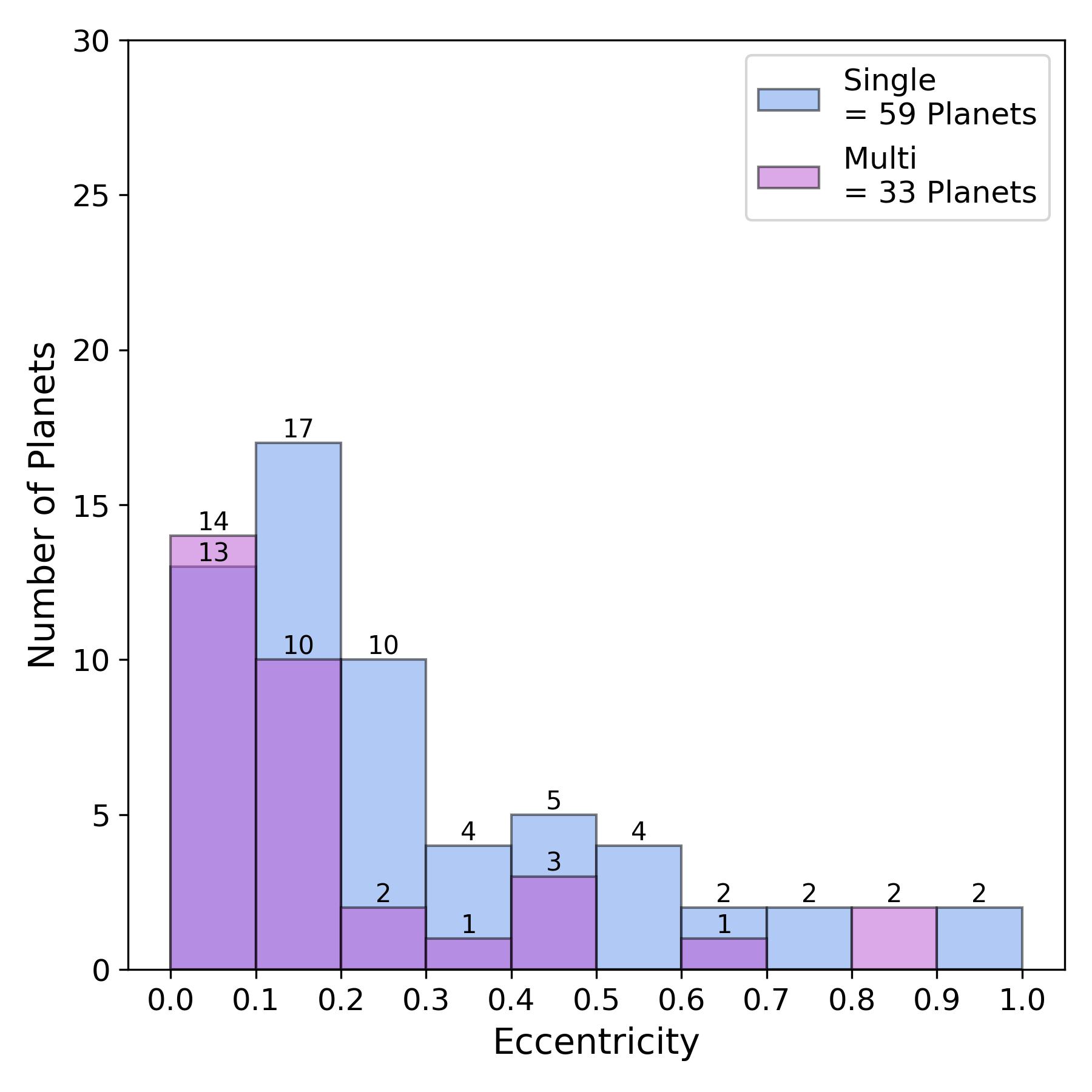} 
      \caption{Eccentricity distribution of single (blue) vs. multiple (magenta) planet systems in our sample.} 
      \label{fig:ecc_hist_single_vs_multi}
    \end{figure}

    Finally, we compare the eccentricity distributions for single ($N=1$) vs. multiple planet ($N>1$) systems (see Fig. \ref{fig:ecc_hist_single_vs_multi}).
    We find a $p$-value < 0.05 from the K-S test, indicating that these two sub-samples of exoplanets are statistically different from each other, and that they could be driven by different evolution paths and processes.
    
    We found that planet multiplicity plays a significant role in the eccentricity distribution of TLGs, where single vs. multi-planet systems produced statistically different distributions (see Fig. \ref{fig:ecc_hist_single_vs_multi}), indicating that these populations could be driven by different evolution paths and processes. 
    The majority of planets with $e > 0.4$ are single-planet systems, while multi-planet systems tend preferentially towards lower eccentricities.
    These findings are compatible with previous predictions, where high eccentricity planets are more likely to be single planets and multi-planet systems are expected produce lower eccentricities \citep{kane2012eccdistkepler, xie2016ecckepler}.
    Finally, it is worth noting that single-planet and 2-planet systems are more likely to have longer period companions that have not yet been discovered due to observation or detection limits, so this could also play a role in the inferred correlation between planet multiplicity and eccentricity.
    Thus, we stress the importance of longer RV follow-up campaigns for the single systems in our sample, where outer planets could still be missing. 
    Another possibility for single-systems with no detected planetary companions could be that the companion could have already been ejected from the system, but there is no way to test or trace this scenario observationally.

    The anti-correlation of planet multiplicity and eccentricity has been predicted by past works, where the eccentricity origin of exoplanets is assumed to be predominantly caused by planet–planet interactions \citep{davies2014protostars}, and was empirically tested in subsequent studies using large samples of RV-detected exoplanets. 
    It is important to note that our sample size (92 planets) is much smaller compared to other studies of planet multiplicity due to our focus on TLGs.
    While the results may not be statistically significant due to the small sample size, it can be useful to compare them with previous works and see whether we find similar correlations.
    \cite{limbach2015multiplicity} found a power law trend of $e(N) = 0.58 \times N^{-1.2}$ for $N>2$ (using the median eccentricities) for a sample of 403 RV-detected exoplanets with non-zero eccentricities.
    \cite{zinzi2017multiplicity} found a power law trend of $e(N) = 0.63 \times N^{-1.02}$ for $N>1$ (using the weighted average eccentricities) for a sample of 258 planets around stars with $T_{\text{eff,$\star$}}$ between 2600$-$7920 K.
    \cite{bachmoller2021multiplicity} found a power law trend of $e(N) = 0.43 \times N^{-0.93}$ for $N>1$ (using the mean eccentricities) for a sample of 1171 exoplanets, the largest statistical sample explored so far.
    We note that our $\alpha$ value (0.35) is close to the one found by \cite{bachmoller2021multiplicity}, while our $\beta$ value (-1.13) falls exactly in between what is found by \cite{zinzi2017multiplicity} and \cite{limbach2015multiplicity}.
    Additionally, we find a similar pattern to \cite{bachmoller2021multiplicity}, where the observed $N=1$ systems (single-planets) have mean and median eccentricities much lower than what is expected from the best-fit models. 
    \cite{bachmoller2021multiplicity} concluded as a result of this that the single-planet sub-sample is likely affected by different evolutionary pathways in comparison with their multi-planet counterparts, and this might also be the case for the TLGs in our sample. 
    Similar to these previous studies, we found an anti-correlation trend between planet multiplicity and eccentricity for our sample of TLGs, further supporting the conclusion that the eccentricity origin of our targets is most likely dominated by planet–planet interactions.

    \subsection{Planet Equilibrium Temperature}\label{subsec:teq_pl}
    
    We test the dependence of the eccentricity distribution of TLGs on the equilibrium temperature of the planet $T_{\text{eq}}$.
    We calculate $T_{\text{eq}}$ (assuming a Bond albedo of 0) using Equation (3) from \cite{kempton2018teq}, as follows,
    
    \begin{equation}\label{eq:Teq_pl}
    T_{\text{eq}} = T_{\text{eff,$\star$}} ~ \sqrt{\frac{R_{\star}}{a}} ~ \Bigg(\frac{1}{4}\Bigg)^{1/4}
    \end{equation}
    
    \noindent where $T_{\text{eff,$\star$}}$ is the effective temperature of the host star, $R_{\star}$ is the stellar radius and $a$ is the semi-major axis of the planet. 
    When calculating $T_{\text{eq}}$, we use the \texttt{jaxstar}-derived $T_{\text{eff,$\star$}}$ and $R_{\star}$ values for each system.
    For homogeneity, we use Kepler's Third Law to calculate $a$, as follows,
    
    \begin{equation}\label{eq:sma}
    a \propto \Big(M_{\star} P^{2}\Big)^{1/3}
    \end{equation}
    
    \noindent where $M_{\star}$ is the \texttt{jaxstar}-derived stellar mass and $P$ is the orbital period of the planet. 
    
\begin{figure}
      \centering
      \includegraphics[width=\linewidth]
      {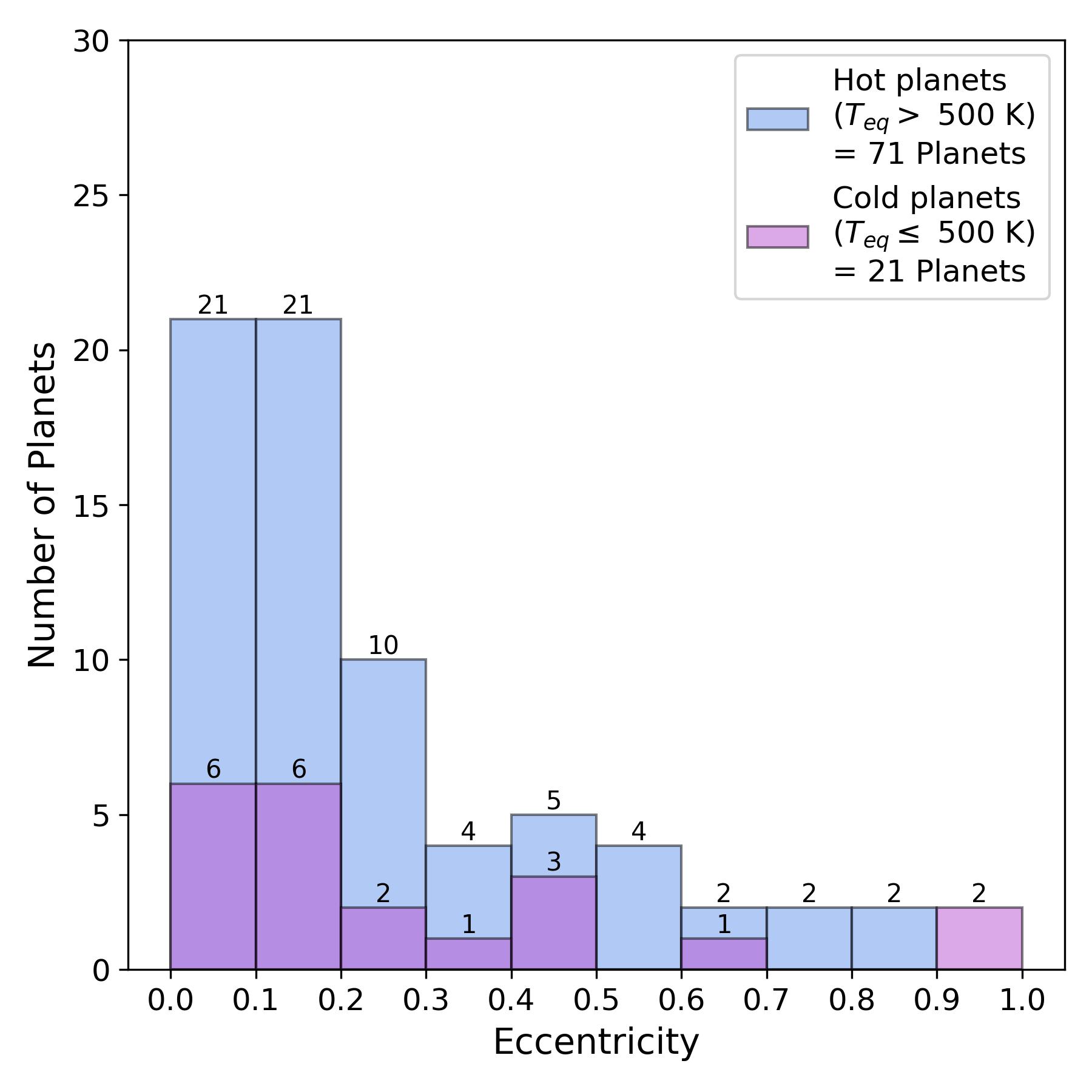} 
      \caption{Eccentricity distribution of systems with hot (blue) vs. cold (magenta) planet equilibrium temperatures in our sample, using 500 K as the cut-off threshold between the two sub-samples.} 
      \label{fig:ecc_hist_cold_vs_hot}
    \end{figure}

    To check for statistical trends using the K-S test, we split our sample based on whether planets had a cold or hot equilibrium temperature, using $T_{\text{eq}}$ = 500 K as the cut-off threshold between the two sub-samples.
    We adopted this threshold because planets with $T_{\text{eq}} \leq$ 500 K were categorized as long-period cold ice/gas giants by previous studies \citep{konatham2020atmospheric, russell2023equilibrium}, while planets with $T_{\text{eq}}$ > 500 K were found to be close-in planets and hot jupiters \citep{konatham2020atmospheric}.
    Fig. \ref{fig:ecc_hist_cold_vs_hot} shows the eccentricity distribution of planet systems with hot (blue) vs. cold (magenta) equilibrium temperatures in our sample.
    Both results from the K-S test and the Spearman $\rho$ test have $p$-values > 0.05, indicating that there is no statistical correlation between $T_{\text{eq}}$ and eccentricity for TLGs. 
    As a sanity check, we tested whether there was any eccentricity correlation with the effective stellar temperature $T_{\text{eff,$\star$}}$.
    As before, the $p$-values are larger than 0.05 from both the K-S test and the Spearman $\rho$ test, indicating that there is no statistical correlation between $T_{\text{eff,$\star$}}$ and eccentricity.

    While previous statistical studies of eccentricity distributions did not directly test the dependence of eccentricity on the planet equilibrium temperature, other independent studies exploring the habitability of exoplanets have looked into this extensively.
    \cite{dressing2010habitable} showed that planets with higher eccentricities could remain habitable at much larger semi-major axes.
    They also reported that larger eccentricities caused planet temperatures to experience increased regional and seasonal variability, leading to a more gradual transition between habitable and non-habitable zones.
    Another study by \cite{linsenmeier2015climate} reported that eccentric orbits typically resulted in two stable climate states, although the range was more limited.
    Perhaps the most relevant dependence found with respect to our study was reported by \cite{mendez2017teqecc}, who found that for a constant albedo, a planet's average equilibrium temperature is expected to decrease as the eccentricity increases.
    We found no such correlation in our sample of TLGs, but this trend could yet be verified by extending the sample size to include non-transiting, long-period giant planets.

    \subsection{Tidal Dissipation}\label{subsec:tidaldissipation}

    We explore the relationship between the eccentricity distribution of TLGs and the tidal dissipation effects on the planet.

        \subsubsection{Circularization Timescale}\label{subsubsec:tcirc}
    
        We calculate the tidal circularization timescale  $t_{\text{circ}}$ using Equation (1) from \cite{trilling2000tidalcirc} or Equation (4) from \cite{jackson2008tidalcirc}, as follows,
        
        \begin{equation}\label{eq:tcirc}
        t_{\text{circ}} = Q_{\text{p}} ~ \sqrt{\frac{a}{G M_{\star}^{3}}} ~ \Bigg(\frac{4}{63}\Bigg) ~ \Bigg(\frac{M_{\text{p}}}{R_{\text{p}}^{5}}\Bigg) ~ a^{6}
        \end{equation}
        
        \noindent where $Q_{\text{p}}$ is the modified tidal quality factor of the planet, $a$ is the orbital semi-major axis of the planet, $G$ is the gravitational constant, $M_{\star}$ is the stellar mass, $M_{\text{p}}$ is the planet mass, and $R_{\text{p}}$ is the planet radius. 
        $Q_{\text{p}}$ encompasses the proper quality factor $Q$, as well as the planet Love number $k_2$, given by $Q_{\text{p}} = (3/2) \times (Q/k_2)$.
        While Equation (\ref{eq:tcirc}) is not strictly appropriate for higher eccentricity regimes (e.g. $e \geq 0.2$), it can be suitable to use when trying to estimate the current eccentricity damping timescale.
        The large uncertainties in the constant time lag approach and the constant tidal quality factor approach, coupled with our ignorance about the initial conditions of the system, make a detailed computation of the evolution of the eccentricity very difficult. 
        In such cases, Equation (\ref{eq:tcirc}) can provide a good estimate of the present-day eccentricity damping timescale.
        In general, the energy change of a planet due to strong tidal interactions with the host star can be estimated as an impulse approximation, which is strongly dependent on the pericentric distance $q$. 
        When $q$ is large enough (e.g. $q$ > 0.05 - 0.1 au), the tidal interactions are too inefficient to damp the planet's eccentricity and semi-major axis. 
        In such cases, the timescale ratio $\tau$ (the ratio between the stellar age and the planet tidal circularization timescale; see Section \ref{subsubsec:tratio}) of high-$e$ planets must be small, as shown in Figs. \ref{fig:t_ratio_vs_ecc_min_max_shaded} and \ref{fig:t_ratio_vs_ecc_varying_Qp_shaded} as well.
        
        When calculating $t_{\text{circ}}$, we use the \texttt{jaxstar}-derived stellar mass for $M_{\star}$ (see Section \ref{sec:jaxstar} for more details). 
        $a$ is calculated using Kepler's Third Law, as shown in Equation (\ref{eq:sma}).
        Finally, we tested two different $Q_{\text{p}}$ cases when calculating $t_{\text{circ}}$: a constant $Q_{\text{p}}$ value and a varying $Q_{\text{p}}$ value (see Section \ref{subsubsec:tidalquality} for more details).

        \subsubsection{Tidal Quality Factor}\label{subsubsec:tidalquality}
        
\begin{figure}
          \centering
          \includegraphics[width=\linewidth]
          {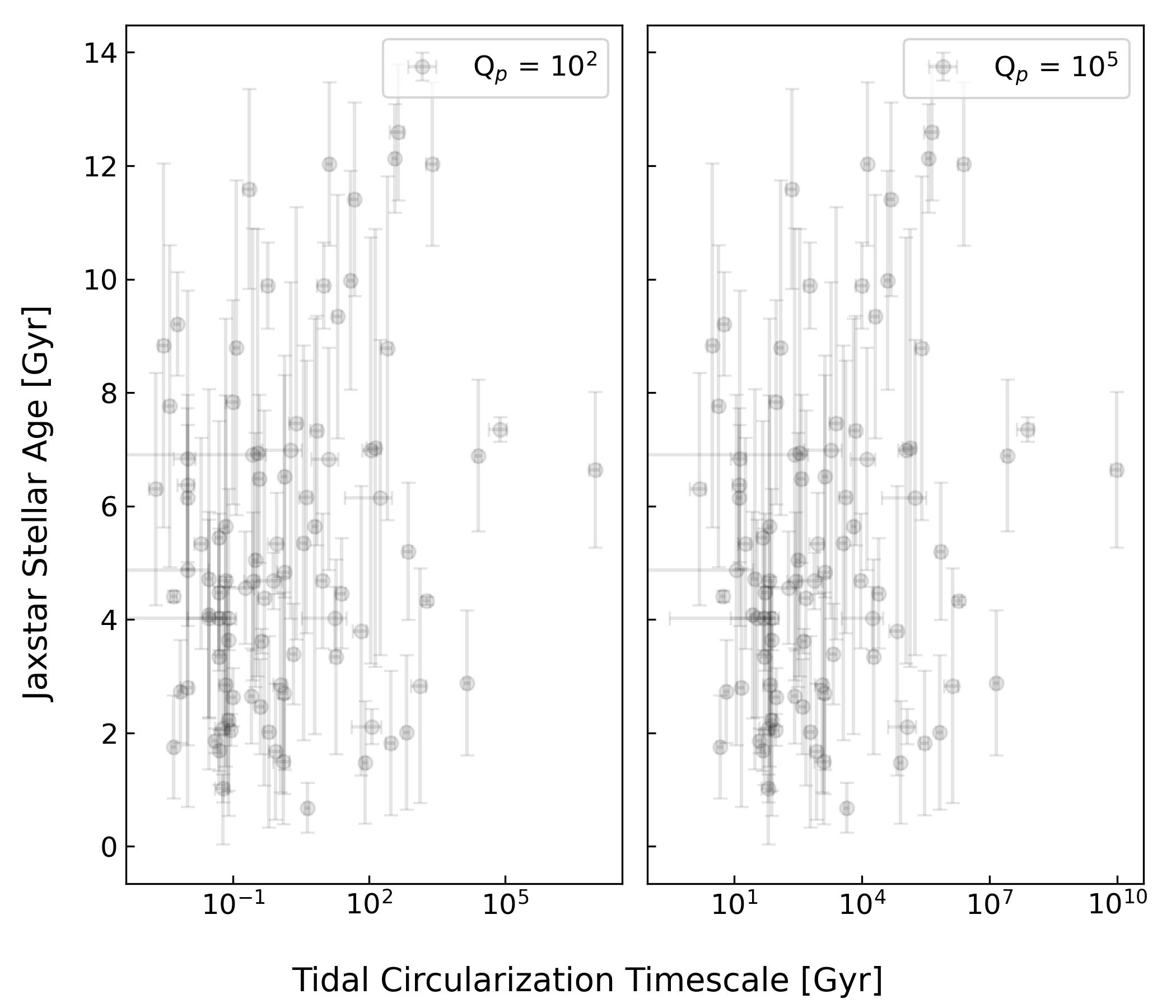} 
          \caption{Tidal circularization timescale vs. stellar age $t_{\text{age}}$ for targets in our sample, assuming constant $Q_{\text{p}}$ scale factors of 10$^{2}$ (left) and 10$^{5}$ (right) for all planets in our sample.} 
          \label{fig:jaxstar_age_vs_tidal_circ_min_max}
        \end{figure}

        We tested two different $Q_{\text{p}}$ cases when calculating the tidal circularization timescale $t_{\text{circ}}$: a constant $Q_{\text{p}}$ value and a varying $Q_{\text{p}}$ value. 
        For the constant $Q_{\text{p}}$ scale factor, we tested 4 different values across the entire target list: 10$^{2}$, 10$^{3}$, 10$^{4}$, and 10$^{5}$.
        Fig. \ref{fig:jaxstar_age_vs_tidal_circ_min_max} shows the tidal circularization timescale $t_{\text{circ}}$ vs. stellar age $t_{\text{age}}$ for all planets in our sample, assuming constant $Q_{\text{p}}$ scale factors for all planets in our sample.
    
        The rheological response and viscosity of planetary materials contribute to the efficiency of tidal dissipation. The tidal quality factor should be related to the internal composition and the thermal state of a planet.
        To determine how to appropriately estimate varying $Q_{\text{p}}$ values for our target list, we use the results of our internal composition modeling, as described in Section \ref{sec:internal}.
        If the interior of a planet can have a significant H/He envelope, we adopt high $Q_{\text{p}}$ values for gas giants. For bare rocky planets or rocky planets with < a few \% H/He envelopes, we use low $Q_{\text{p}}$ values similar to those for terrestrial planets. A planet that may contain a water mantle is assumed to have intermediate $Q_{\text{p}}$ values, such as Uranus and Neptune. Following the concept above, planets in groups (i) and (ii) are expected to have high $Q_{\text{p}}$ values ($\sim 10^{4}$-$10^{5}$), while planets in groups (iii) and (iv) are expected to have intermediate $Q_{\text{p}}$ values ($\sim$ 10$^{2}$-10$^{3}$), and group (v) planets are expected to have low $Q_{\text{p}}$ values ($\sim$ 10-100).
        Based on this, we divided our sample into 3 categories depending on their radius, and we assigned different $Q_{\text{p}}$ values to the planets in each category as follows:
        \begin{enumerate}
            \item High $Q_{\text{p}}$ ($10^{5}$): planets with $R_{\text{p}}$ > 6 $R{_{\oplus}}$
            \item Intermediate $Q_{\text{p}}$ (10$^{3}$): planets with 3 $R{_{\oplus}}$ < $R_{\text{p}}$
            < 6 $R{_{\oplus}}$ 
            \item Low $Q_{\text{p}}$ (10$^{2}$): planets with $R_{\text{p}}$ < 3 $R{_{\oplus}}$
        \end{enumerate}
    
\begin{figure}
          \centering
          \includegraphics[width=\linewidth]
          {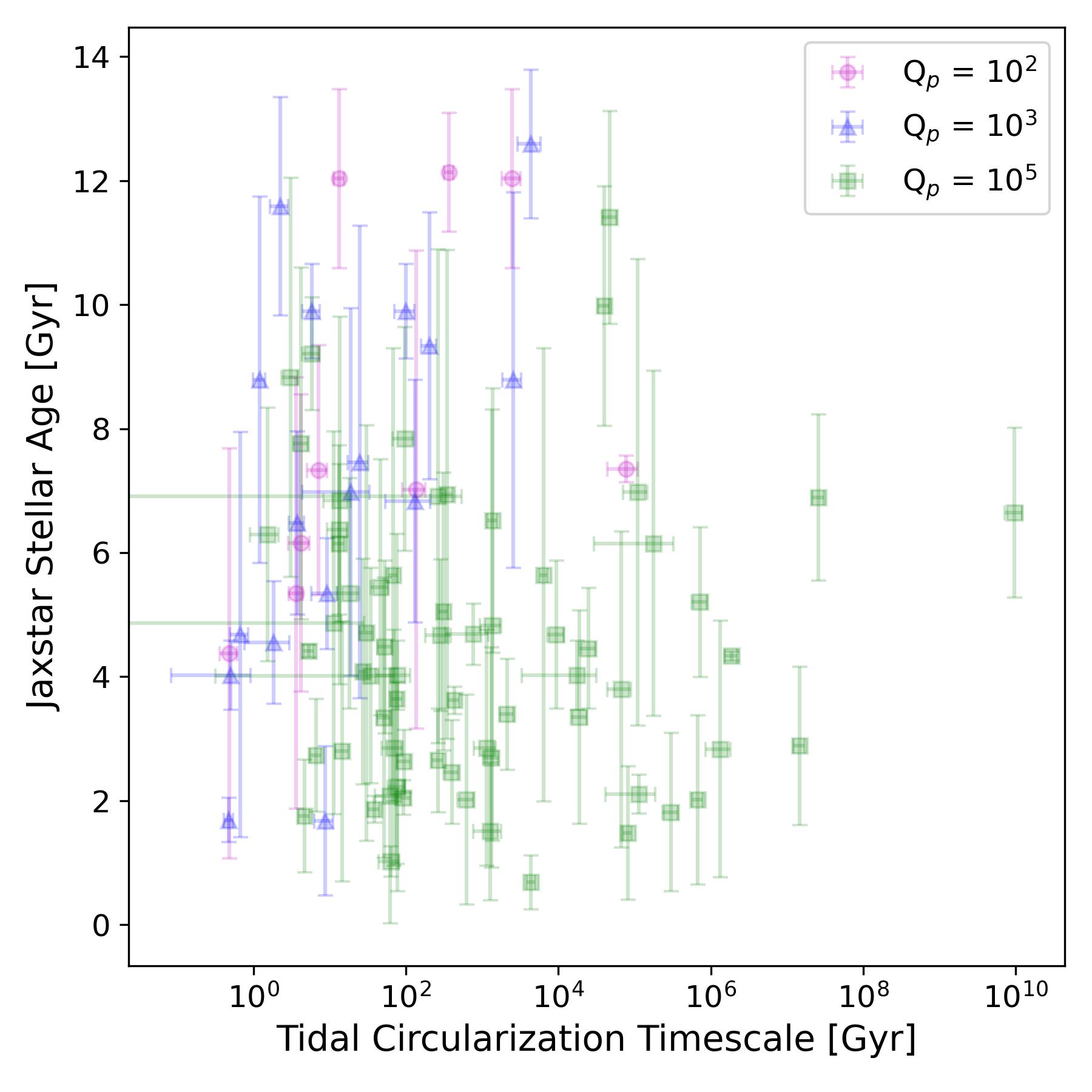} 
          \caption{Tidal circularization timescale vs. stellar age $t_{\text{age}}$ for targets in our sample, assuming varying $Q_{\text{p}}$ scale factors based on the planet's radius. We highlight the $Q_{\text{p}}$ values of 10$^{2}$, 10$^{3}$ and 10$^{5}$ in circle magenta points, triangle blue points and square green points, respectively.} 
          \label{fig:jaxstar_age_vs_tidal_circ_varying_Qp}
        \end{figure}

        Fig. \ref{fig:jaxstar_age_vs_tidal_circ_varying_Qp} shows $t_{\text{circ}}$ vs. $t_{\text{age}}$ again, but assuming varying $Q_{\text{p}}$ scale factors for the planets in our sample using the parametrization based on the planet's radius.
    
        \subsubsection{Short vs. Long Tidal Timescale Planets}\label{subsubsec:shortvslongtcirc}
    
        To compare the eccentricity distribution of planets with short vs. long tidal circularization timescales, we define the timescale ratio $\tau$ (the ratio between the stellar age $t_{\text{age}}$ and the planet's tidal circularization timescale $t_{\text{circ}}$) as follows,
    
        \begin{equation}\label{eq:tratio}
        \tau = \frac{t_{\text{age}}}{t_{\text{circ}}}
        \end{equation}
        
        For $t_{\text{age}}$, we use the \texttt{jaxstar}-derived stellar ages, as described in Section \ref{sec:jaxstar}.
        We use the timescale ratio to split our sample using a cut-off threshold of $\tau = 0.9$ when checking for statistical correlations using the K-S test.
        At $\tau = 1$, $t_{\text{circ}}$ is as long as the age of the system $t_{\text{age}}$. 
        Since $t_{\text{circ}}$ is still very close to the age of the system at $\tau = 0.9$, we chose to use 0.9 as the cut-off instead of 1. 
        When $\tau > 0.9$, $t_{\text{age}}$ is longer than $t_{\text{circ}}$, meaning that the planet is likely to circularize during its lifetime.
        When $\tau \leq 0.9$, $t_{\text{circ}}$ is longer than $t_{\text{age}}$, indicating that the planet is not likely to circularize during its lifetime and will retain its present eccentricity.
    
\begin{figure}
        \centering
                \includegraphics[width=\linewidth]{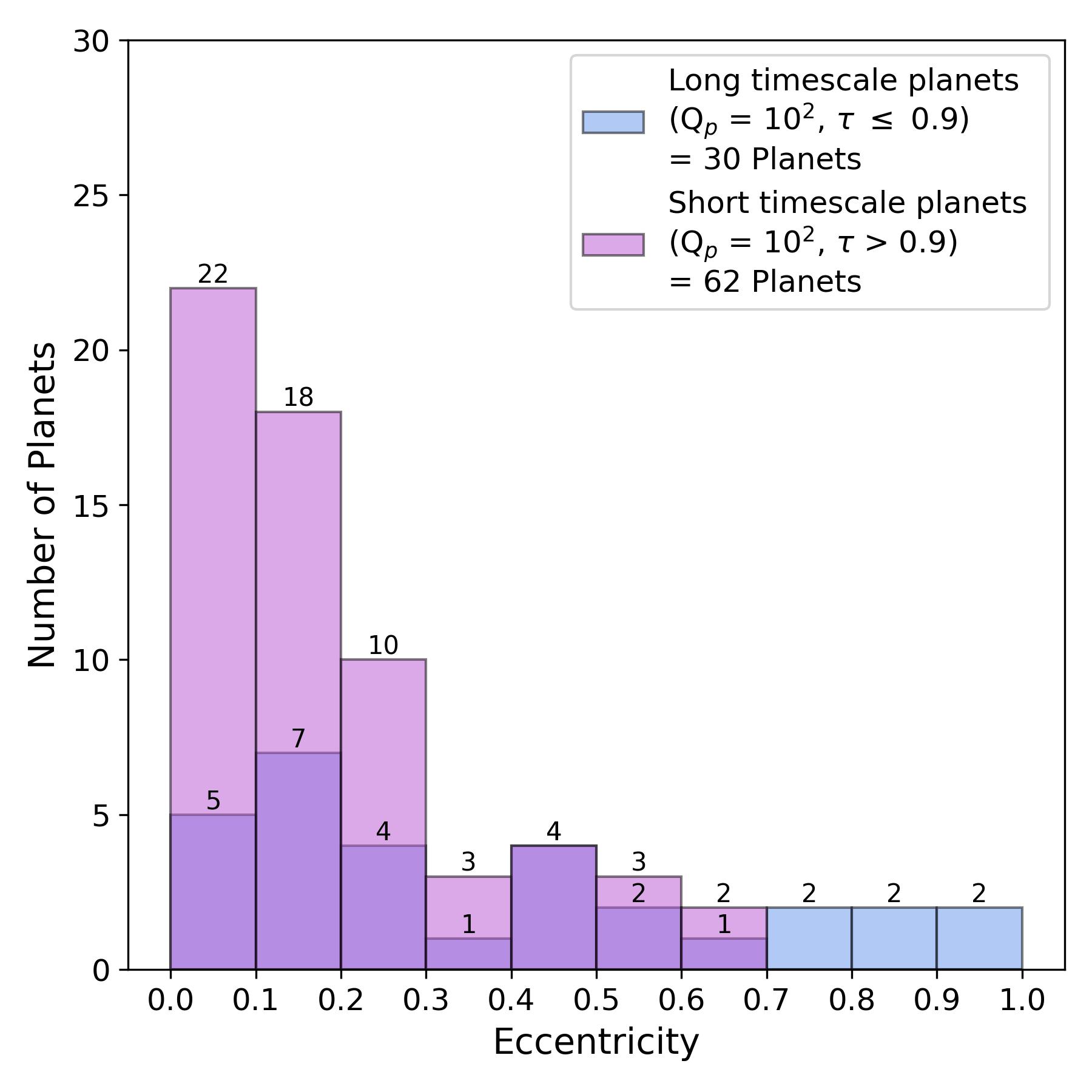}
                \includegraphics[width=\linewidth]{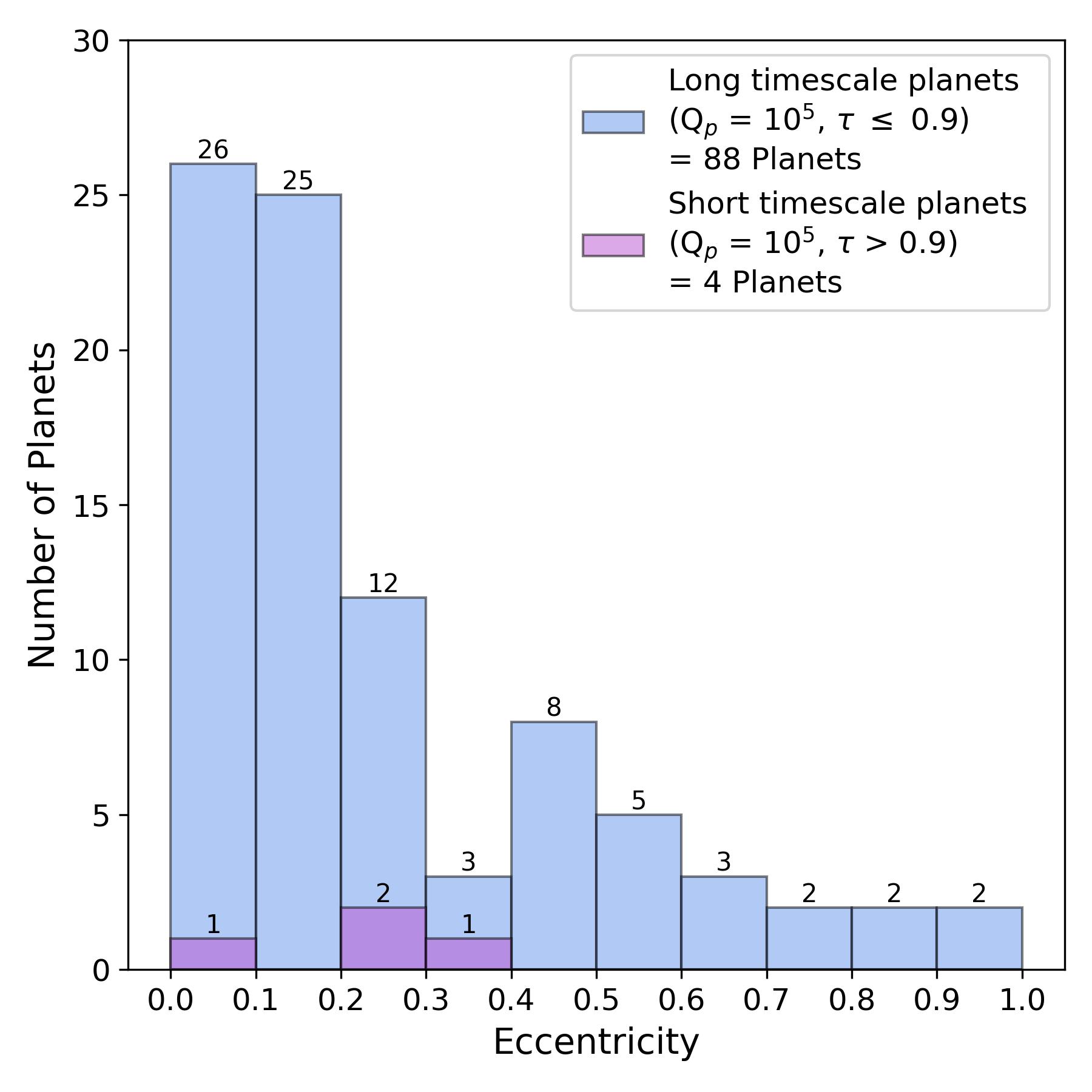}
        \caption{Eccentricity distribution of planets with long (blue) vs. short (magenta) tidal circularization timescales with respect to their host-star age (using a cut-off of $\tau = 0.9$), assuming constant $Q_{\text{p}}$ scale factors of 10$^{2}$ (top) and 10$^{5}$ (bottom).}
          \label{fig:ecc_hist_tcirc}
        \end{figure}

        We computed the eccentricity distribution of planets with long (blue) vs. short (magenta) tidal circularization timescales with respect to their host-star age (using a cut-off of $\tau = 0.9$), assuming 4 constant $Q_{\text{p}}$ scale factors (10$^{2}$, 10$^{3}$, 10$^{4}$, and 10$^{5}$).
        We performed statistical tests to check the significance of the two distributions for each constant $Q_{\text{p}}$ value.
        Both results from the K-S test and the Spearman $\rho$ test have $p$-values > 0.05 for $Q_{\text{p}}$ scale factors of 10$^{3}$, 10$^{4}$, and 10$^{5}$, indicating that there is no statistical significance between the short vs. long timescale distributions for these cases.
        For the for $Q_{\text{p}}$ = 10$^{2}$ case, the K-S test has a $p$-value < 0.05, while the Spearman $\rho$ test yields a $p$-value > 0.05. 
        Fig. \ref{fig:ecc_hist_tcirc} shows the eccentricity distributions for the lowest (10$^{2}$) and highest (10$^{5}$) $Q_{\text{p}}$ scale factor values.
    
\begin{figure}
            \centering
            \includegraphics[width=\linewidth]{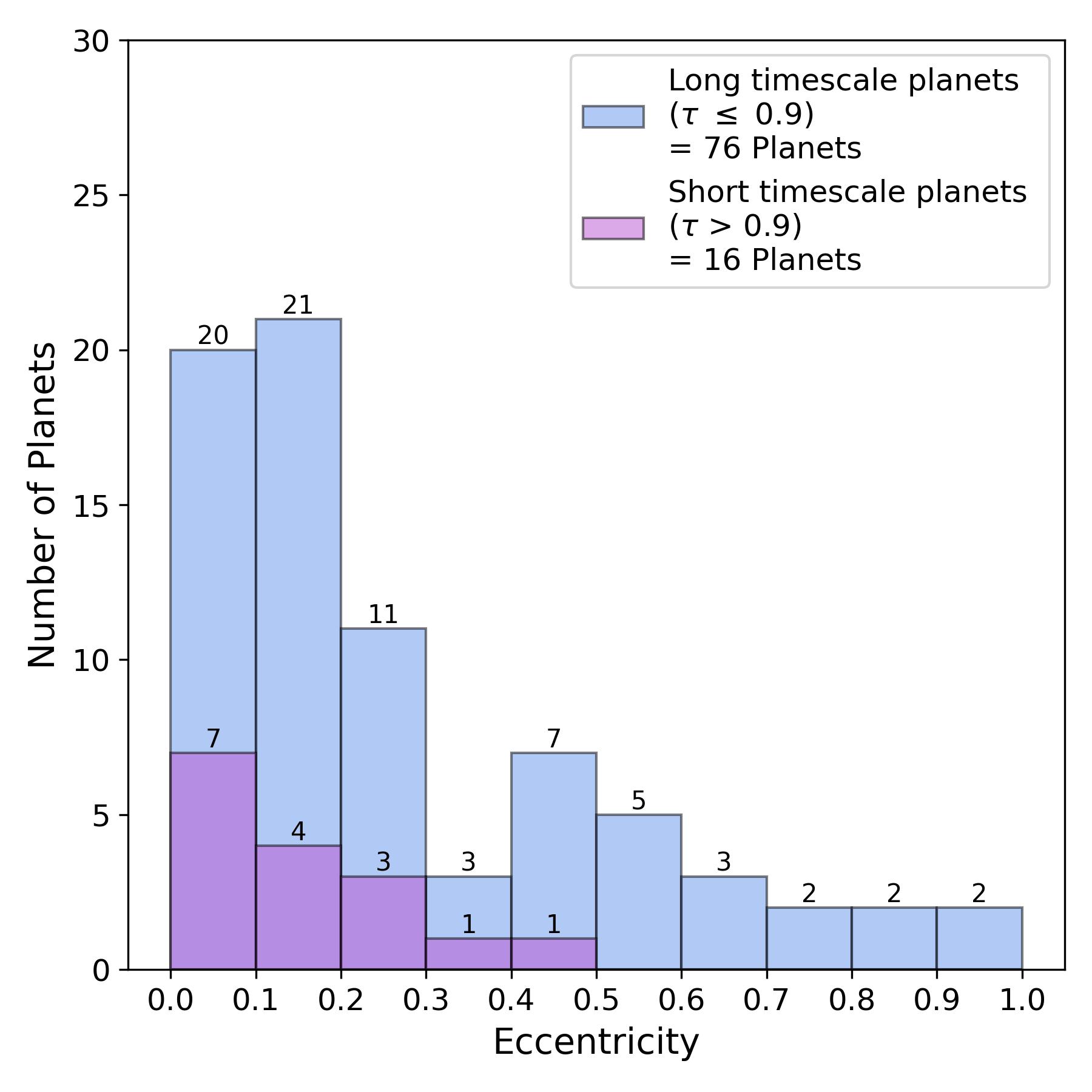}
            \caption{Eccentricity distribution of planets with long (blue) vs. short (magenta) tidal circularization timescales with respect to their host-star age (using a cut-off threshold of $\tau = 0.9$) assuming varying $Q_{\text{p}}$ scale factors based on the planet composition.}
            \label{fig:ecc_hist_short_vs_long_tcirc_varying_Qp}
        \end{figure}

        Fig. \ref{fig:ecc_hist_short_vs_long_tcirc_varying_Qp} shows the eccentricity distribution of planets with long (blue) vs. short (magenta) tidal circularization timescales with respect to their host-star age (using a threshold of $\tau = 0.9$) assuming 3 varying $Q_{\text{p}}$ scale factors based on the planet composition.
        We performed K-S tests to check the significance of the two distributions, and find a $p$-value of 0.31 ($p > 0.05$), indicating that there is no statistical significance between the short vs. long timescale distributions.
        
        \subsubsection{Timescale Ratio}\label{subsubsec:tratio}
        
        Since we can only broadly estimate what the appropriate $Q_{\text{p}}$ values are for our TLG sample, we choose to focus on the general trend observed in the timescale ratio $\tau$ with respect to eccentricity, so that our findings will not be dependent on which $Q_{\text{p}}$ values are assumed.
    
\begin{figure}
          \centering
          \includegraphics[width=\linewidth]
          {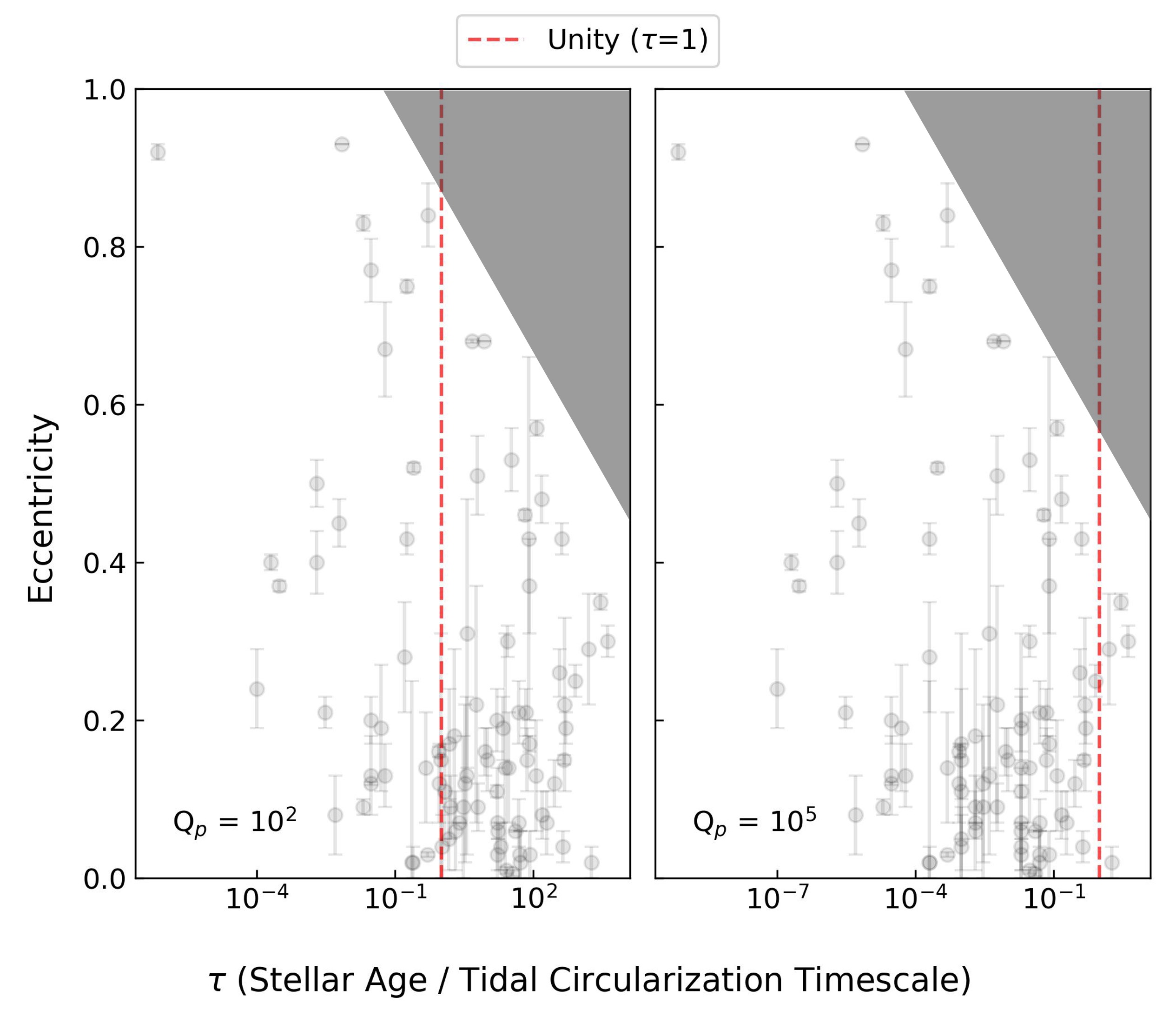} 
          \caption{Timescale ratio $\tau$ vs. eccentricity $e$ for targets in our sample, assuming constant $Q_{\text{p}}$ scale factors. For reference, we also show the unity of $\tau$, where $t_{\text{age}}$ = $t_{\text{circ}}$, using a red-dashed line. The gray shaded areas mark the "forbidden" zone, where planets at higher eccentricities are not found beyond a given $\tau$.} 
          \label{fig:t_ratio_vs_ecc_min_max_shaded}
        \end{figure}

\begin{figure}
          \centering
          \includegraphics[width=\linewidth]
          {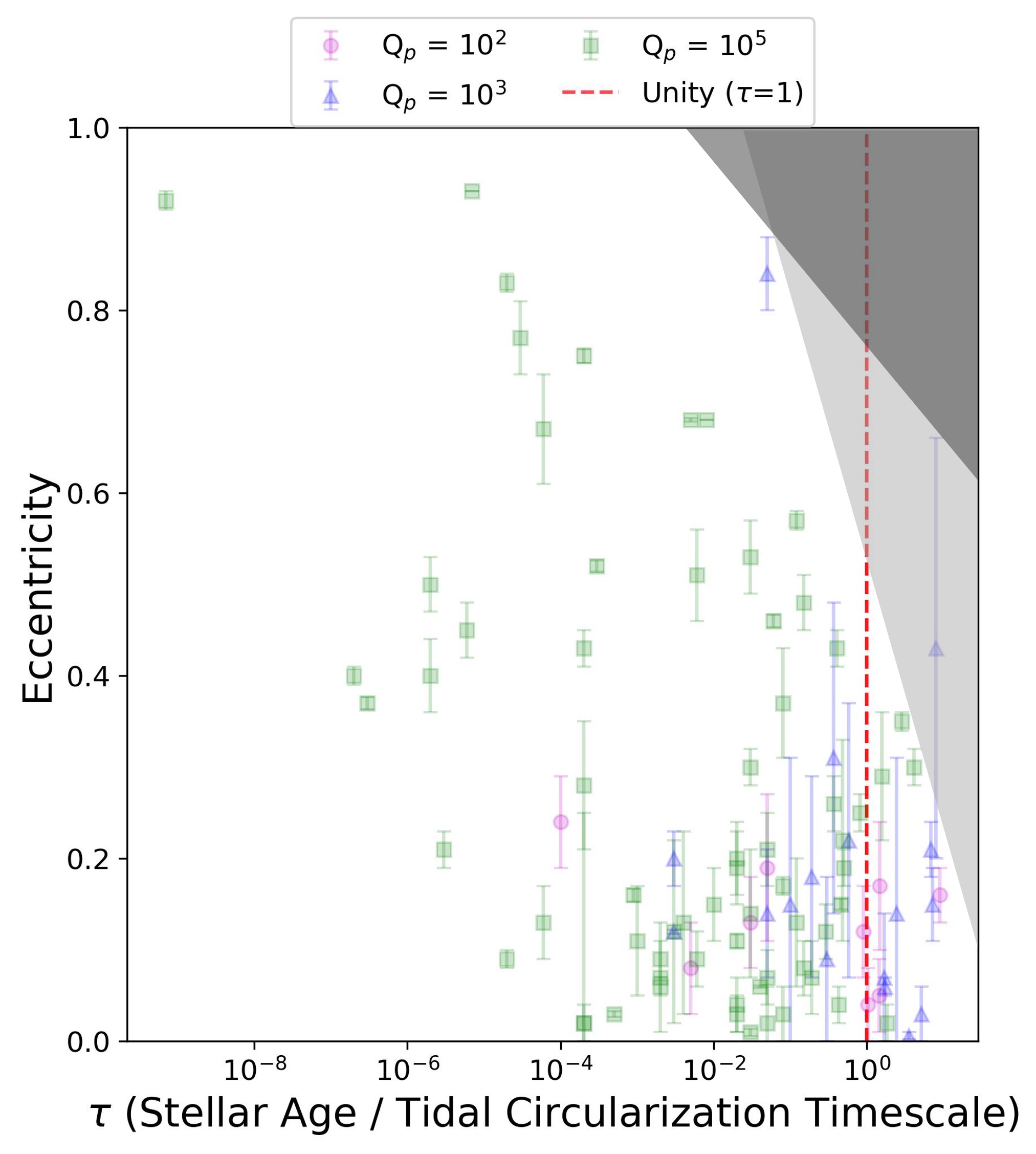} 
          \caption{Timescale ratio $\tau$ vs. eccentricity $e$ for targets in our sample, assuming varying $Q_{\text{p}}$ scale factors. For reference, we also show the unity of $\tau$, where $t_{\text{age}}$ = $t_{\text{circ}}$, using a red-dashed line. The dark-gray shaded region marks the "forbidden" zone, where planets at higher eccentricities are not found beyond a given $\tau$. The light-gray shaded region shows the "forbidden" zone after excluding Kepler-89 c (TIC 273231214), which has large uncertainties on its eccentricity measurement.} 
          \label{fig:t_ratio_vs_ecc_varying_Qp_shaded}
        \end{figure}

        Fig. \ref{fig:t_ratio_vs_ecc_min_max_shaded} shows the timescale ratio $\tau$ vs. eccentricity $e$ for targets in our sample, assuming constant $Q_{\text{p}}$ scale factors, while Fig. \ref{fig:t_ratio_vs_ecc_varying_Qp_shaded} shows the same plot but for varying $Q_{\text{p}}$ scale factors.
        We find a "forbidden" zone, marked by the gray shaded areas in the plots.
        
        In the constant $Q_{\text{p}}$ case, we find that there are no planets with eccentricities of $e \geq 0.4$ beyond a given $\tau$, regardless of the $Q_{\text{p}}$ scale factor used.
        In the varying $Q_{\text{p}}$ case, we highlight 2 different "forbidden" zones.
        The first zone is marked by dark-gray shaded region in the plot, where there are no planets with eccentricities of $e > 0.6$ beyond a given $\tau$.
        The second zone is marked by the light-gray shaded region after excluding one target, Kepler-89 c (TIC 273231214), which has large uncertainties on its eccentricity measurement.
        When excluding Kepler-89 c, this second "forbidden" zone extends down to $e \sim 0.1$, beyond which there are no planets.
    
        \subsubsection{Implications}\label{subsubsec:tidaldissipation_discussion}

        There is still much work to be done when it comes to understanding how best to quantify $Q_{\text{p}}$ based on planet properties (e.g. density, equilibrium temperature, etc.) \citep{mathis2018tidesreview}.
        Some studies have tried to estimate $Q_{\text{p}}$ using tidal dissipation measurements of Solar System planets and moons \citep{lainey2009jupitertides, fuller2024saturntides}, which could in theory help us apply them to exoplanets \citep{dhouib2023tidesgiants, lazovik2024tidesgiants}.
        Giant planets are more likely to have higher $Q_{\text{p}}$ values (e.g. 10$^{4}$-10$^{5}$) \citep{mahmud2023tidalq}, but there is still no direct way to map $Q_{\text{p}}$ quantitatively (or with a good degree of certainty) to exoplanets based on their observed properties.
    
        Since we are targeting TLGs in our sample, and $Q_{\text{p}}$ values of $\sim 10^5$ are more likely for Jovian planets, we take a closer look at the $Q_{\text{p}}$ = 10$^{5}$ for the constant $Q_{\text{p}}$ case in Fig. \ref{fig:t_ratio_vs_ecc_min_max_shaded} (the right-most panel).
        Almost all planets are located to the left of the red-dashed line (the unity of the timescale ratio $\tau$, where $t_{\text{age}}$ = $t_{\text{circ}}$), meaning that the majority of planets in our sample are not likely to circularize during their lifetime (see the eccentricity distribution in Fig. \ref{fig:ecc_hist_tcirc} for the same constant $Q_{\text{p}}$ value, where only 4 planets have expected tidal timescales shorter than their host-star age).
        
        In the varying $Q_{\text{p}}$ case, our internal composition modeling revealed that the majority of our targets are Jovian planets, where high $Q_{\text{p}}$ values ($\sim 10^5$) are more likely, which is consistent with what is expected for TLGs.
        In this case as well, Fig. \ref{fig:t_ratio_vs_ecc_varying_Qp_shaded} shows that almost all planets are located to the left of the red-dashed line (the unity of the timescale ratio $\tau$, where $t_{\text{age}}$ = $t_{\text{circ}}$). 
        This indicates that we might be probing the intrinsic, initial state of the system's eccentricity and that the majority of planets in our sample are not likely to circularize during their lifetime (see also the eccentricity distribution in Figs. \ref{fig:ecc_hist_tcirc} and \ref{fig:ecc_hist_short_vs_long_tcirc_varying_Qp}, where only 4 and 16 planets have expected tidal timescales shorter than their host-star age, respectively).
        
        Our findings from both $Q_{\text{p}}$ cases indicate that the eccentricity distributions of our TLG sample might be a reflection of their primordial state, without having experienced any significant tidal dissipation.
        The lack of strong tidal dissipation effects further reinforces our finding that the orbital eccentricity of TLGs does not seem to correlate with the presence of stellar companions (as discussed in Section \ref{subsec:stellarcomp}).
        
        Lastly, we find an interesting "forbidden" zone, marked by the gray shaded areas in Figs. \ref{fig:t_ratio_vs_ecc_min_max_shaded} (constant $Q_{\text{p}}$) and \ref{fig:t_ratio_vs_ecc_varying_Qp_shaded} (varying $Q_{\text{p}}$), where planets at higher eccentricities are not found beyond a given $\tau$, regardless of the $Q_{\text{p}}$ scale factor used.
        The forbidden zone could indicate that the intrinsic eccentricity distribution in our sample is expected to be at $\tau$ values much shorter than 1. 
        For instance, in Fig. \ref{fig:t_ratio_vs_ecc_varying_Qp_shaded}, the intrinsic eccentricity distribution is likely only visible in $\tau$ smaller than where the forbidden region is (e.g. less than $\sim 10^{-3}$).
        
        There are several possible scenarios to explain the origin of the forbidden zone.
        The first scenario could be that planets with high eccentricities are no longer able to form at larger $\tau$ values.  
        Without a (sub-)stellar companion or other massive planets in the vicinity, highly eccentric planets would not be able to form in the forbidden region of the system due to the absence of strong perturbers.
        However, planets may well form with high eccentricity in that region, but they could be subject to a fast eccentricity decay, removing them from the forbidden zone.
        An alternative scenario could be related to strong tidal interactions between the planet and its host star.
        Planets with high eccentricities should have experienced strong tidal interactions if their pericentric distance $q$ is small enough (e.g. $q$ < 0.05 au) during their closest approach to their host star, after which their eccentricity and semi-major axis decrease rapidly due to small tidal damping timescales. 
        In the case of tidal interactions, the eccentricity damping timescale is comparable to the orbital decay timescale, meaning that planets will have moved closer to their host star in this scenario. 
        Currently, we see that these planets have low eccentricities at small semi-major axes, and planets in the forbidden zone may have disappeared.
        As a result, we would not expect to see any planets with high eccentricities and large $\tau$ values.
        The habitat of planets in a circular orbit around 0.1 AU may originate from tidal circularization of planets in the forbidden region.
        
        Interestingly, we found that high-$e$ planets (most of which are gas giants with high $Q_{\text{p}}$ values) tend to have lower $\tau$ values, suggesting that the orbits of low-$e$ planets are susceptible to tidal circulation. 
        \cite{udry2007exostats} reports that for periods in the range $P \sim$ 10-30 days, which is considered to be distinctly outside the circularization period by tidal interaction with the star, there are a few systems with very low eccentricities.
        As part of our sample selection criteria, we placed a period cut of $P$ > 10 days to exclude planets that would be severely affected by tides, but the low-$e$ planets in our sample still appear to be sensitive to it.

    \subsection{Stellar Metallicity}\label{subsec:stellarmetallicity}
    
\begin{figure}
      \centering
      \includegraphics[width=\linewidth]
      {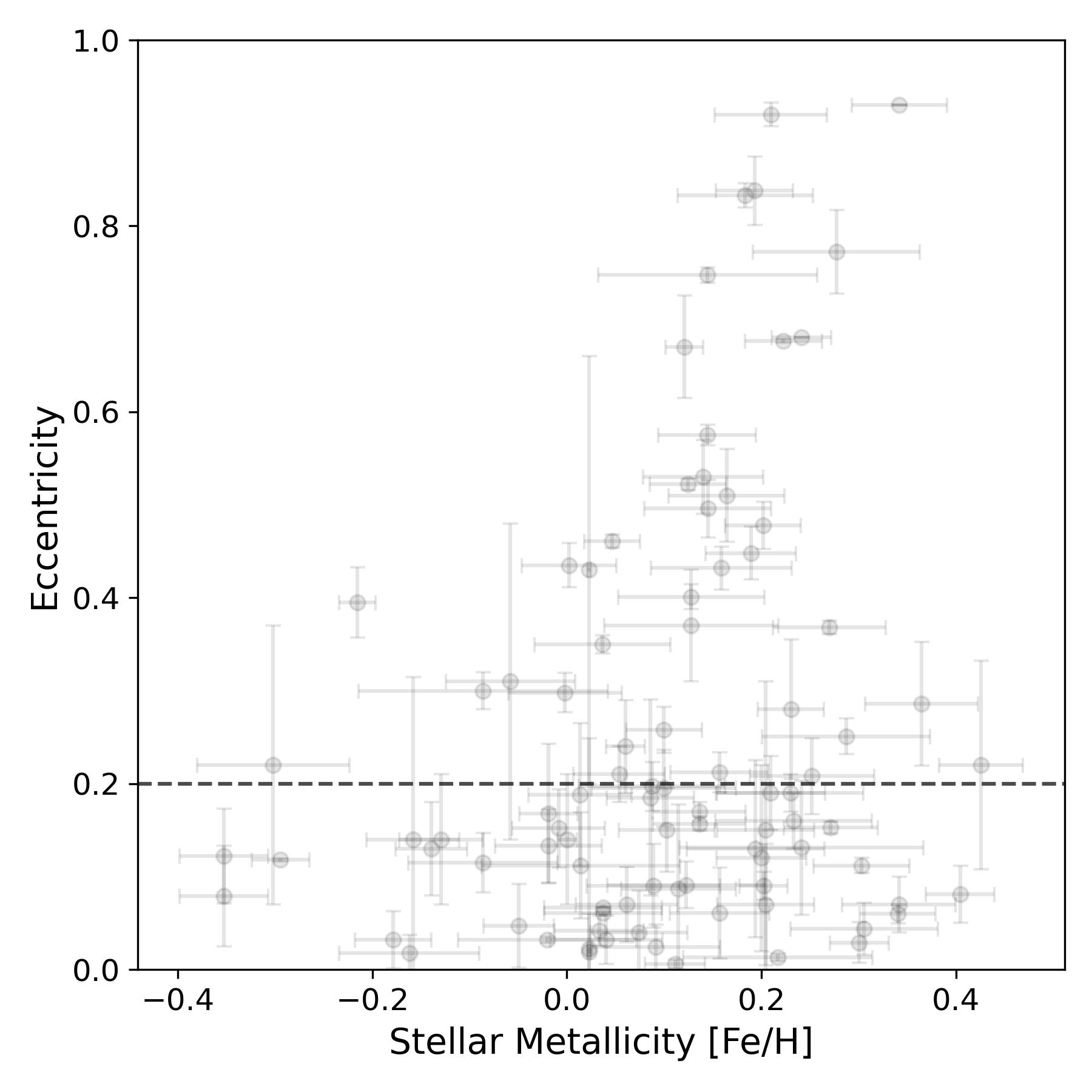} 
      \caption{Stellar Metallicity [Fe/H] vs. Eccentricity for all 92 planets in our sample.} 
      \label{fig:jaxstar_feh_vs_e}
    \end{figure}

    We explore the stellar metallicity correlation to the planet eccentricity for our sample of TLGs.
    Fig. \ref{fig:jaxstar_feh_vs_e} shows the stellar metallicity [Fe/H] vs. eccentricity for all 92 targets in our sample.
    A visible trend can be seen in the data, where host-star metallicities are homogeneously distributed for planets with $e \leq 0.4$, but narrow down significantly at higher eccentricities around $\sim 0.1 < \text{[Fe/H]} < 0.3$.
    Figure \ref{fig:jaxstar_feh_vs_e} clearly indicates that highly eccentric planets are only found around metal-rich stars with [Fe/H] > 0.1.
    
\begin{figure}
      \centering
      \includegraphics[width=\linewidth]
      {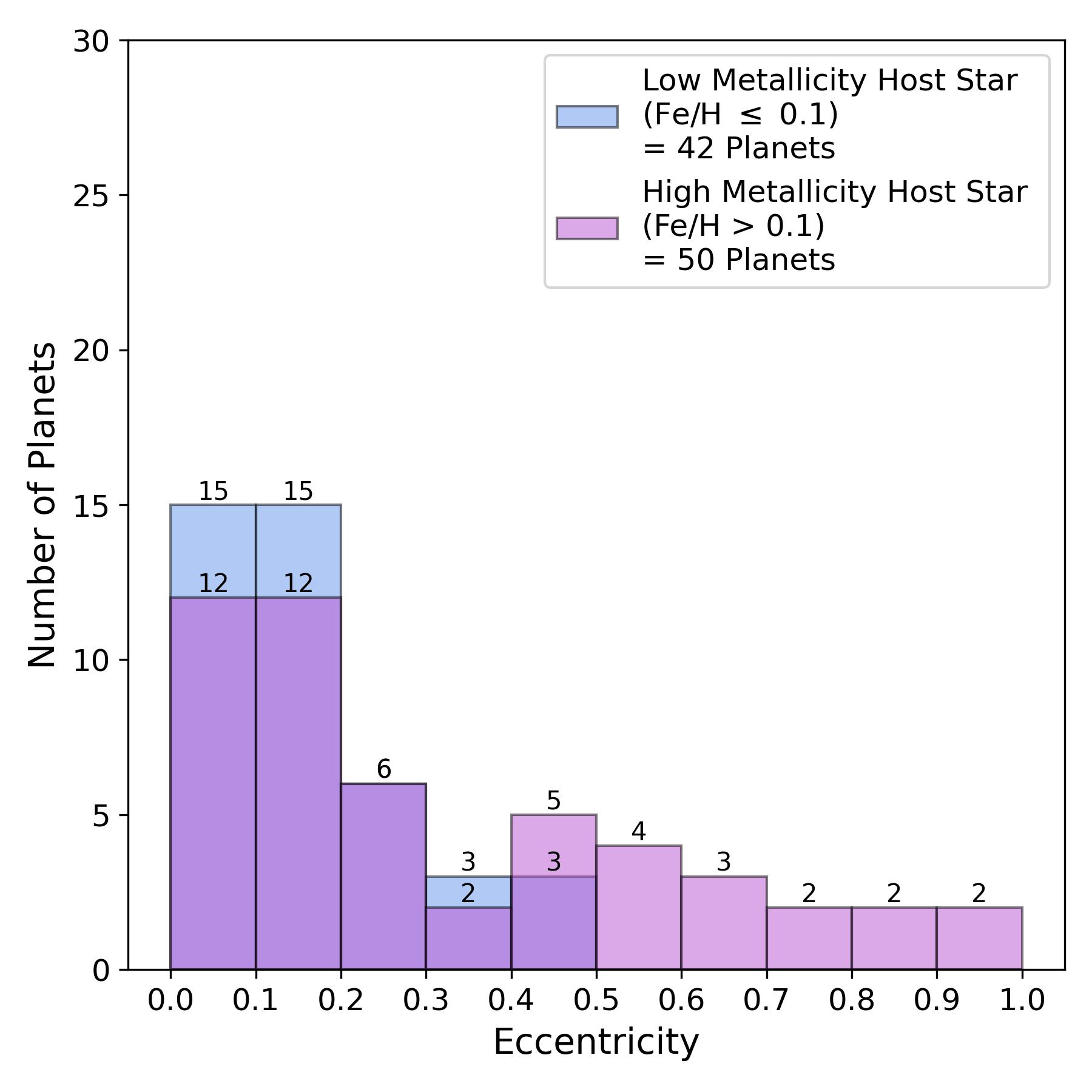} 
      \caption{Eccentricity distribution of planet systems with lower (blue) vs. higher (magenta) metallicity host-stars in our sample, using [Fe/H] = 0.1 as the cut-off threshold between the two sub-samples.} 
      \label{fig:ecc_hist_metal_poor_vs_metal_rich}
    \end{figure}

    To test the statistical significance of this correlation using the K-S test, we split our sample based on whether planets belong to lower or higher metallicity host-stars, using [Fe/H] = 0.1 as the cut-off threshold between the two sub-samples.
    This threshold was chosen based on the median metallicity of our sample.
    We compare the eccentricity distributions of the two sub-samples in Fig. \ref{fig:ecc_hist_metal_poor_vs_metal_rich}. 
    We find that there are no planets beyond $e > 0.4$ in the lower metallicity host-star sample, while the higher metallicity host-star sample includes planets across the entire range of eccentricities.
    Both results from the K-S test and the Spearman $\rho$ test have $p$-values < 0.05, indicating that there is metallicity-dependence on eccentricity.
    
    Previous studies show that metal-rich stars have a higher probability of harboring a giant planet than their lower metallicity counterparts \citep{udry2007exostats}, highlighting the crucial role metallicity plays in the formation and evolution of giant planets.
    In fact, many papers have found that stars hosting giant planets were observed to be more metal-rich \citep{gonzalez1997metallicity, fuhrmann1997solartype, santos2001metalrich, fischer2005metallicitycorr, bond2006abundance}.
    Additionally, the frequency of planets (regardless of eccentricity) is found to be higher around metal-rich host stars, and the dependence of planet abundance on stellar metallicity has been widely known and extensively studied \citep{bond2006abundance, adibekyan2019heavymetal}.
    \cite{bond2006abundance} found that planet-hosting stars were observed to have significantly higher mean metallicities ([Fe/H] = 0.06 $\pm$ 0.03 dex) compared to non-planet-hosting stars ([Fe/H] = $-$0.09 $\pm$ 0.01 dex).
    Interestingly, \cite{adibekyan2019heavymetal} mentions the necessity of separating hot and cold jupiters when deriving metallicity correlations. 
    This is because the orbital period of giant planets also correlates with metallicity, and longer-period cold giant planets tend to orbit more metal-poor stars than their shorter-period hot giant planet counterparts.
    
    In our study, we found a clear correlation between stellar metallicity on the eccentricity distribution of TLGs, a dependence which was only previously proven for close-in giant planets \citep{dawson2013planetscat}.
    Relations between the stellar metallicity and other planetary orbital parameters have also been previously explored (e.g. $e$, $a$, and $M_{\text{p}} \sin i$), but no significant trends have been found \citep{laws2003parentstars, santos2003statproperties, fischer2005metallicitycorr}.
    While \cite{santos2003statproperties} reported that there were no statistical correlations between planet eccentricity and the stellar metallicity, their findings suggested that low [Fe/H] stars appear to only host planets with intermediate eccentricities, and more eccentric planets were only found around stars with higher metallicities.
    These findings are in agreement with our results, which show a statistically significant correlation between metallicity and eccentricity.
    Planets belonging to lower metallicity host-stars ([Fe/H] $\leq 0.1$) in our sample only had eccentricities up to $e = 0.4$, indicating that these stars can only host low to moderate eccentricity planets.
    In contrast, planets belonging to higher metallicity host-stars ([Fe/H] $> 0.1$) spanned the entire range of eccentricities ($0 < e < 1$).
    
    A metallicity dependence suggests that more giant planets are expected to form around host-stars with higher [Fe/H], with some planets having more enhanced eccentricities and others being entirely ejected from the system. 
    Similarly, since low-[Fe/H] stars form fewer large planets, low-[Fe/H] stars could be hosting more low-$e$ planets due to the lack of giant planets that could cause eccentricity excitation via planet-planet scattering.    
    For low-$e$ gas giants found around low-[Fe/H] stars, planet-planet scattering is unlikely to occur because gas giants form preferentially in metal-rich environments by the core accretion model. 
    Several directly imaged gas giants on wide orbits, which are difficult for the core accretion model to form, have been found around metal-poor stars, such as the multi-gas giant system around HR 8799. 
    A low [Fe/H] could be favorable for the disk instability scenario.
    
\begin{figure}
      \centering
      \includegraphics[width=\linewidth]
      {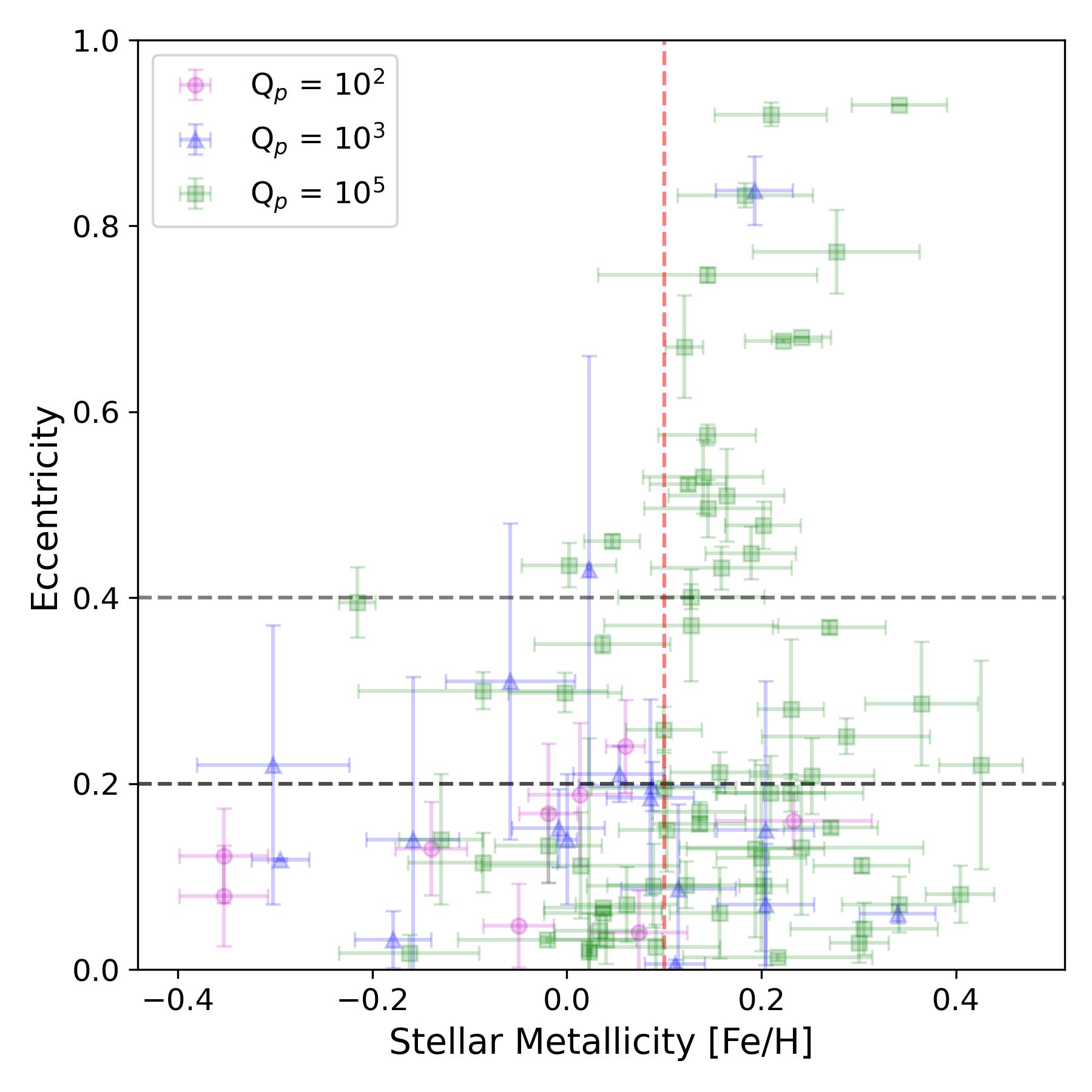} 
      \caption{Stellar Metallicity [Fe/H] vs. Eccentricity for all 92 planets in our sample. The data points are colored according to the varying $Q_{\text{p}}$ values estimated from the internal composition modeling of the planets. We highlight the $Q_{\text{p}}$ values of 10$^{2}$, 10$^{3}$ and 10$^{5}$ in circle magenta points, triangle blue points and square green points, respectively.} 
      \label{fig:jaxstar_feh_vs_e_colorQp}
    \end{figure}

    Fig. \ref{fig:jaxstar_feh_vs_e_colorQp} shows a plot of [Fe/H] vs. $e$, colored according to the varying $Q_{\text{p}}$ values estimated from the internal composition modeling of the planets.
    For clarity, we display two horizontal gray dashed lines at $e=0.2$ (below which planets are broadly considered to have low-$e$ or non-eccentric orbits) and $e=0.4$ (the eccentricity up to which disk migration could occur and be a possible source of the planet's eccentricity \citep{bitsch2010orbitalevol, debras2021migration}).
    The vertical dashed red line at [Fe/H] = 0.1 is the cut-off threshold used to split our planet sample between lower-metallicity and higher-metallicity host stars.
    The plot shows that there is a mixture of planet compositions at low [Fe/H] (to the left side of the red dashed line), even up to $e=0.4$. 
    In contrast, gas giants (with the highest $Q_{\text{p}}$ value, colored in green) appear to dominate more at higher [Fe/H] (to the right side of the red dashed line).
    For the subset of low-$e$ gas giants found at lower metallicities, disk instability could be a more plausible scenario for the origin of these systems.
    
\begin{figure}
      \centering
      \includegraphics[width=\linewidth]
      {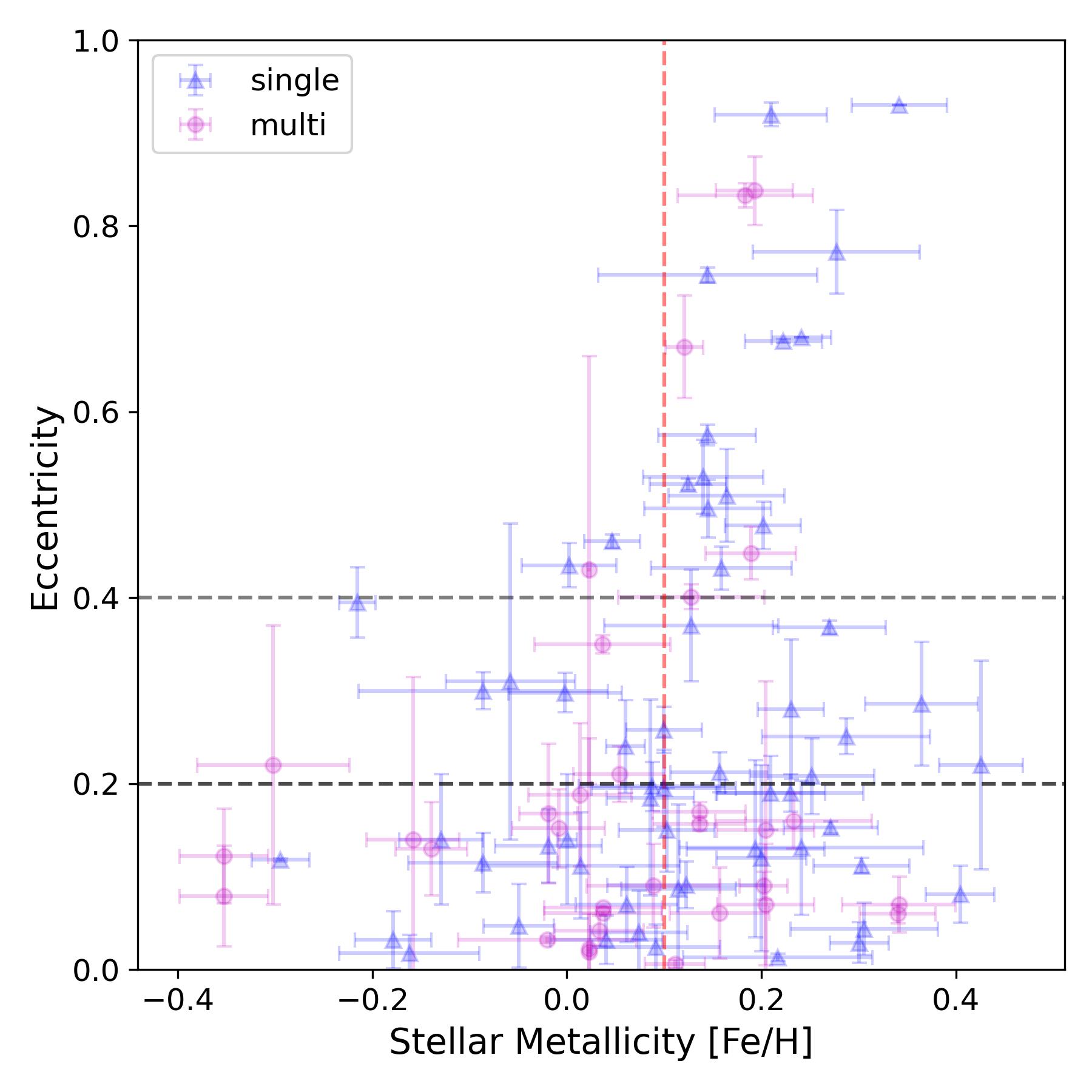} 
      \caption{Stellar Metallicity [Fe/H] vs. Eccentricity for all 92 planets in our sample. The data points are colored according to the planet multiplicity, with single systems and multi-planet systems highlighted in blue triangle points and circle magenta points, respectively.} 
      \label{fig:jaxstar_feh_vs_e_colornplanet}
    \end{figure}

    Additionally, we checked whether there is any visible correlation between planet multiplicity and the presence of low-$e$ planets around low [Fe/H] stars.
    Fig. \ref{fig:jaxstar_feh_vs_e_colornplanet} shows a plot of [Fe/H] vs. $e$, colored according to planet multiplicity. 
    As before, we display two horizontal gray dashed lines at $e=0.2$ and $e=0.4$.
    We find no visible trend for planet multiplicity within the low-$e$ and low-[Fe/H] planet regime.

    \cite{knierim2022originofgiants} explored the origin of warm Jupiters within the context of planet formation theory, where the two leading models are (i) formation at the outer disk followed by migration and (ii) in situ formation. 
    They found that migrating giant planets had 2-14 times higher metallicities than planets that formed in situ, and that the metallicity of migrating planets increased with decreasing planetary mass, but was constant for in situ formation.
    This could mean that disk migration is more favorable for high [Fe/H] giant planets in our sample (up to $e=0.4$) rather than forming in situ and being excited via planet-planet scattering.

\section{Possible Biases}\label{sec:biases}

The statistical study of the eccentricity distribution of long-period giant planets has not been previously well-explored.
This could in part be due to the ongoing challenges with the detection and characterization of long-period planets, in particular with respect to the long radial velocity (RV) coverage required to accurately characterize such systems and determine their eccentricities \citep{lagrange2023radialgiants}.
This becomes especially challenging with the presence of stellar and instrumental noise, which are commonly prevalent in RV data and cause contamination depending on the magnitude of the target and the level of its magnetic activity.

It is important to emphasize that the findings of this paper could be biased due to the small sample size, and the target list is not large enough to make any conclusive statements with regards to the wider sample of long-period giant planets.
In Section \ref{sec:sample}, we described the different criteria we used to select our sample, and our decision to focus on transiting systems ultimately limited the sample size.
Additionally, we did not place any radius cuts, so our target list includes a mixture of different types of planets with varying compositions (see the results of our internal composition modeling in Section \ref{sec:internal}).
Interestingly though, \cite{shen2008eccdist} explored different biases arising from RV surveys in relation to the eccentricity distribution of exoplanets, and found that the detection efficiency only slightly decreased with eccentricity.
They reported that the main source of uncertainty in the eccentricity distribution came from biases in Keplerian fits to data with low RV amplitudes and a limited number of observations, rather than from selection effects.
Thus, long-baseline RV campaigns would be very useful for sampling the eccentricity distribution of long-period giant planets.

Another potential observational bias is the heterogeneity of the high-resolution imaging data used in this sample.
As discussed in Section \ref{subsec:stellarcomp}, different telescopes and imaging techniques were used, which resulted in different detection limits.
Some targets have a combination of imaging observations that are complementary to each other (e.g. AO + Speckle), which make them more reliable when determining the presence of stellar companions.
We also mentioned previously how a small number of targets were missing high-resolution imaging altogether, making the sample $\sim$8\% incomplete.
In the ideal case, in order to have a more homogeneous sample of imaging data, all targets would be observed with high-resolution imaging, with multiple complementary observations utilizing both AO and Speckle techniques, and preferably observed with the same telescopes.

We would also like to highlight that this might not be the ideal target list for this type of study, and follow-up analyses that expand this target list to a larger sample (e.g. including non-transiting systems), or place a more rigorous focus on planet composition (e.g. by introducing strict radius cuts) would be beneficial for future work on the eccentricity origin of long-period giant planets.
Performing a homogeneous re-fitting of the radial velocity data would also be ideal in order to counter any possible biases introduced in the reported planet properties as a result of heterogeneous analyses, but this is beyond the scope of this paper.
Large RV campaigns dedicated to monitoring long-period giant planets could play a key role in the eccentricity distributions of these planets, which have otherwise not been well-explored.
Other approaches suggest utilizing hierarchical Bayesian modeling to infer the true eccentricity distribution rather than using a histogram of estimated eccentricities, which could be beneficial for future large-scale population studies \citep{hogg2010inferring}.

\begin{figure}
    \centering
    \includegraphics[width=\linewidth]{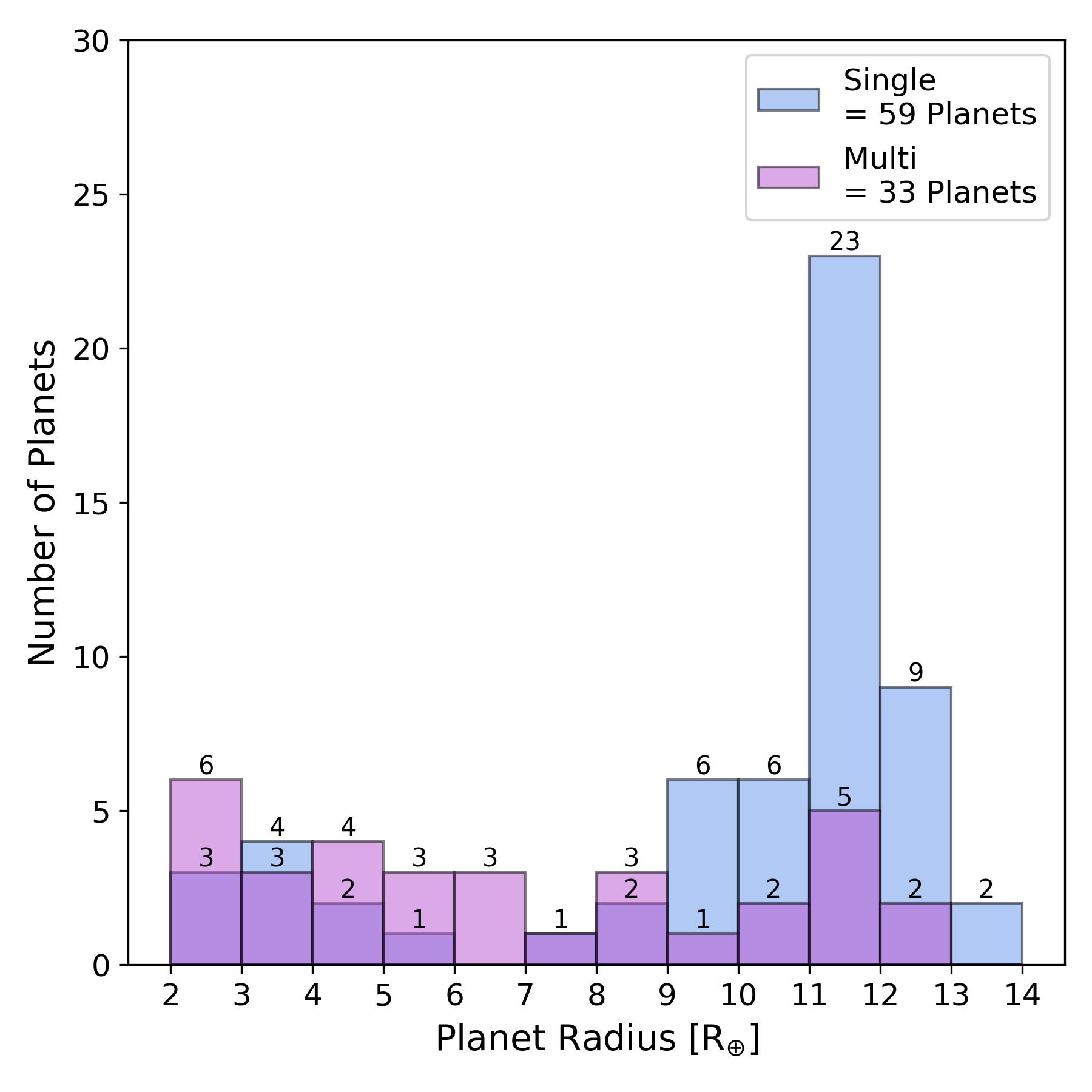}
    \caption{Radius distribution of single (blue) vs. multiple (magenta) planet systems in our sample.}
    \label{fig:radius_hist_single_vs_multi}
\end{figure}

\begin{figure}
    \centering
    \includegraphics[width=\linewidth]{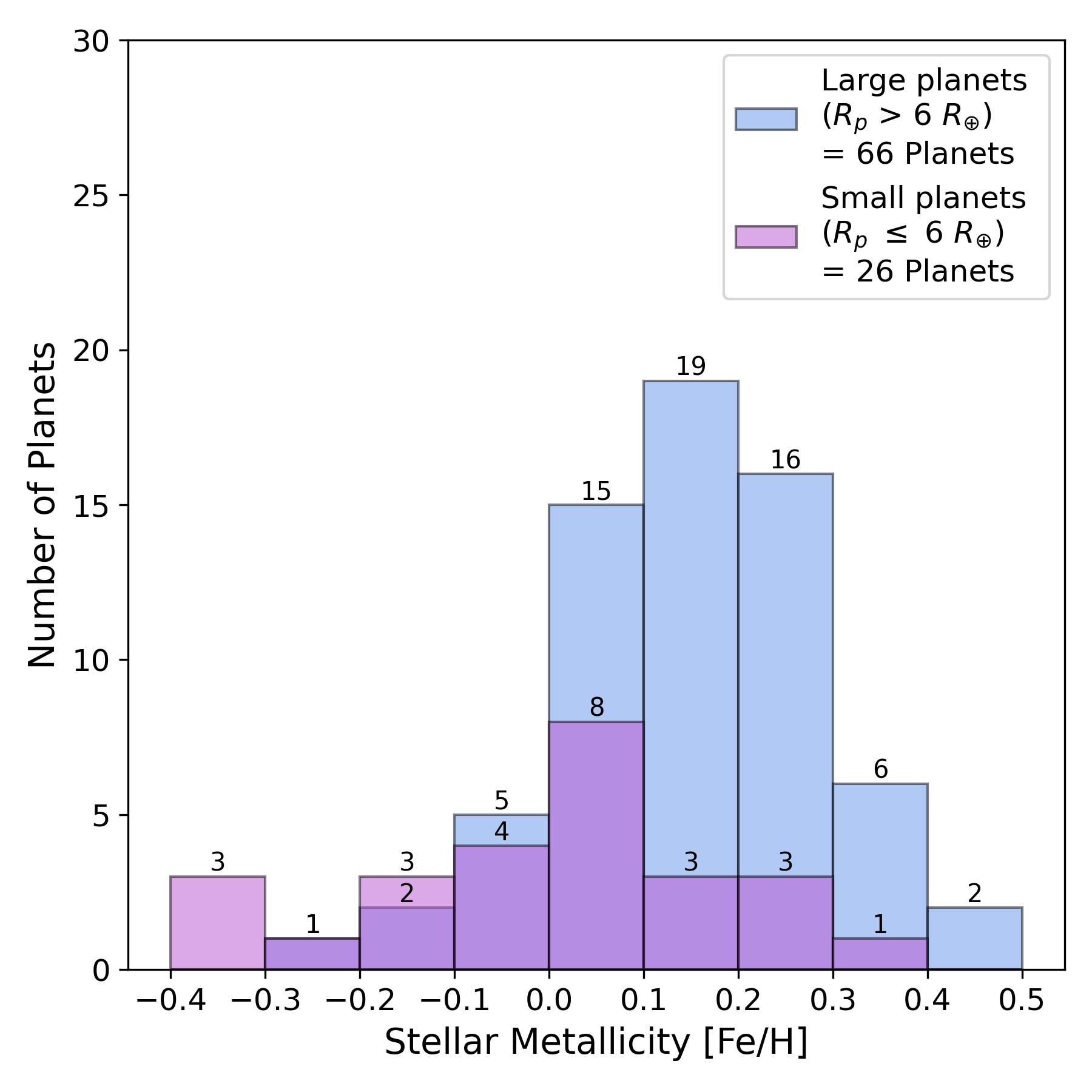}
    \caption{Stellar Metallicity distribution of large (blue) vs. small (magenta) planets, using a cut-off threshold of $R_{\text{p}}$ = 6 $R_{\oplus}$.}
    \label{fig:feh_hist_small_vs_large}
\end{figure}

Finally, while we found correlations for stellar metallicity, planet radius and planet multiplicity with the planet eccentricity, there is a possibility that these parameters are correlated with each other, and this could be contributing to their individual correlations with the planet's eccentricity.
In other words, there could be correlations between these parameters that are independent of eccentricity.
To test whether this is the case, we performed K-S tests between these three parameters.
We found a strong dependence of planet radius on planet multiplicity ($p$-value = 0.00002), and planet radius on stellar metallicity ($p$-value = 0.0007), while no correlations were found between planet multiplicity and stellar metallicity ($p$-value = 0.12).
Fig. \ref{fig:radius_hist_single_vs_multi} shows the radius distribution for single vs. multi-planet systems.
Interestingly, we can see a bimodal radius distribution for single-planet systems, with a gap between $R_{\text{p}}$ = 6-7 $R_{\oplus}$. 
Additionally, the majority of these planets appear to be clustered towards larger radii ($R_{\text{p}} > 7 ~ R_{\oplus}$).
In contrast, the multi-planet systems show a more flat radius distribution, with no gap separating small vs. large planets.
This suggests that single-systems could be more sensitive to formation processes that produce smaller planets compared to their larger counterparts.
However, we highlight that this could also be a result of observational bias since larger planets are easier to detect, coupled with the fact that many single-systems might not be followed up with studies that would detect possible outer companions if any exist.
For eccentricity distributions of small planets ($R_{\text{p}}$ = 1-6 $R_{\oplus}$), \cite{vaneylen2019eccdist} found that the planet radius did not play any significant role for single vs. multi-planet systems, indicating that multiplicity does not depend on the planet size when the small vs. large planet populations are separated.
Fig. \ref{fig:feh_hist_small_vs_large} shows the stellar metallicity distribution as a function of planet radius.
Larger planets ($R_{\text{p}} > 6 ~ R_{\oplus}$) are found to be more abundant at higher metallicities, which is consistent with theoretical predictions since giant planets are expected to form more around metal-rich stars.
Since stellar metallicity and planet multiplicity were not found to be correlated with each other, they are likely independently correlated with eccentricity. 
However, we found a strong radius correlation with other parameters, since both metallicity and multiplicity were sensitive to the planet size.
As a result, the inter-dependencies between radius and these other parameters could partially be responsible for its mutual correlation with the eccentricity distributions of TLGs.

\section{Summary and Conclusions}\label{sec:summary}

Eccentric giant planets are predicted to have acquired their eccentricity through two major mechanisms: the Kozai-Lidov effect or planet-planet scattering, but it is usually difficult to distinguish between the two mechanisms and determine the true eccentricity origin for a given system.
A population study on a statistical level for the observed distribution of such planets could provide better insights into their eccentricity origins and evolution history.
In this work, we focus on a sample of 92 transiting, long-period giant planets (TLGs) as part of an eccentricity distribution study for this planet population in order to understand their eccentricity origin. 
We used archival high-contrast imaging observations, public stellar catalogs, precise \textit{Gaia} astrometry, and the NASA Exoplanet Archive database, to explore the eccentricity distribution correlation with different planet and host-star properties, including: stellar age, stellar metallicity, stellar companion, planet radius, planet multiplicity, planet equilibrium temperature and planet tidal dissipation timescale.
To mitigate biases in heterogeneous model fits, we homogeneously characterized the basic stellar properties for all 86 host-stars in our sample, including stellar age and metallicity.

We found that the planet eccentricity of TLGs depends on stellar metallicity, planet radius and planet multiplicity.
Our findings show that lower-metallicity stars ([Fe/H] $\leq$ 0.1) did not host any planets beyond $e > 0.4$, while higher-metallicity stars hosted planets across the entire eccentricity range ($0 < e < 1$).
Additionally, planet multiplicity played a significant role in the eccentricity distribution of TLGs, where the majority of planets with $e > 0.4$ were single-planet systems, while multi-planet systems leaned preferentially towards lower eccentricities.
The correlation found for planet radius showed that separate planet populations exist within our sample.
This was further supported by the results of our internal composition modeling, which revealed a mixture of planet groups.

We also explored the general trend observed in the timescale ratio $\tau$ with respect to eccentricity, so that our findings were not dependent on which $Q_{\text{p}}$ values were assumed when determining the tidal circularization timescales of our planets.
We found an interesting "forbidden" zone, where there were no planets with high eccentricities and large $\tau$ values.
There are several possibilities for the origin of this zone: 1) planets at higher eccentricities are no longer able to form beyond a given $\tau$, 2) planets form with high eccentricity but are removed from the forbidden zone after experiencing a fast eccentricity decay, or 3) the eccentricities of these planets decrease rapidly due to strong tidal interactions at large $\tau$ values.
The general trend indicates that the eccentricity distributions of our TLG sample might be a reflection of their primordial state, with a lack of strong tidal dissipation effects.
Interestingly, we found no correlation between the eccentricity distribution and the presence of stellar companions, indicating that planet-planet scattering is likely a more dominant mechanism than the Kozai-Lidov effect for TLGs.
This was further supported by an anti-correlation trend found between planet multiplicity and eccentricity, as well as a lack of strong tidal dissipation effects for planets in our sample, which favor planet-planet scattering scenarios for the eccentricity origin.

\section*{Acknowledgements}

We thank the referee for their thoughtful and helpful feedback, which helped improve the paper.
AAQ acknowledges the funding bodies, the UAE Ministry of Presidential Affairs and the UAE Space Agency, for their support through the PhD scholarship. 
DK acknowledges MWGaiaDN, a Horizon Europe Marie Sk\l{}odowska-Curie Actions Doctoral Network funded under grant agreement no. 101072454 and also funded by UK Research and Innovation (EP/X031756/1). 
DK also acknowledges the UK's Science \& Technology Facilities Council (STFC grant ST/W001136/1).
YH was supported by JSPS KAKENHI Grant Number JP24H00017.
I.A.S. acknowledges the support of M.V. Lomonosov Moscow State University Program of Development.
We thank Kento Masuda for the very helpful discussions on estimating the properties of the stellar parameters in this study.
This research has made use of the Exoplanet Follow-up Observation Program (ExoFOP; DOI: 10.26134/ExoFOP5) website, which is operated by the California Institute of Technology, under contract with the National Aeronautics and Space Administration under the Exoplanet Exploration Program. 
This work has also made use of data from the European Space Agency (ESA) mission \textit{Gaia} (\url{https://www.cosmos.esa.int/gaia}), processed by the \textit{Gaia}
Data Processing and Analysis Consortium (DPAC, \url{https://www.cosmos.esa.int/web/gaia/dpac/consortium}). 
Funding for the DPAC has been provided by national institutions, in particular the institutions participating in the \textit{Gaia} Multilateral Agreement.

\section*{Data Availability}

The data tables used and presented in this work are publicly available at CDS via anonymous ftp to \url{https://cdsarc.u-strasbg.fr} (130.79.128.5).
The coding scripts used for the analyses can be made available upon reasonable request.


\bibliographystyle{mnras}
\bibliography{bibliography}



\appendix

\section{NEXA Parameters}\label{sec:nexaparams}

\onecolumn

\renewcommand*{\arraystretch}{1.3}
\begin{longtable}{lllllllllll}
\caption{NEXA planet parameters of our target list that were used in our study.}
\label{table:nexaplanetparams}
\\ \hline
       TIC ID &      Planet Name &          $P$ &  $\sigma_{P}$ &  $R_{\text{p}}$ &  $\sigma_{R_{\text{p}}}$ &  $M_{\text{p}}$ &  $\sigma_{M_{\text{p}}}$ &        $e$ &  $\sigma_{e}$ &  $N$ \\
              &                  &       [days] &        [days] &  [$R_{\oplus}$] &           [$R_{\oplus}$] &  [$M_{\oplus}$] &           [$M_{\oplus}$] &            &               &      \\
\hline
TIC 298663873 &       TOI-2180 b &    260.79 &         0.58 &          11.32 &                    0.23 &         875.62 &                   26.70 &  0.368 &        0.007 &   1 \\
TIC 172370679 &       TOI-1899 b &  29.09031 &      0.00004 &          11.10 &                    0.34 &         212.95 &                   12.71 &   0.04 &         0.03 &   1 \\
TIC 377780790 &      Kepler-10 c &  45.29430 &      0.00005 &           2.36 &                    0.02 &          11.40 &                    1.30 &   0.13 &         0.05 &   3 \\
TIC 432254760 & EPIC 249893012 c &    15.624 &        0.001 &           3.67 &                    0.15 &          14.67 &                    1.86 &   0.07 &         0.06 &   3 \\
 TIC 73717937 &        HATS-17 b &  16.25461 &      0.00007 &           8.71 &                    0.63 &         425.26 &                   20.66 &   0.03 &         0.02 &   1 \\
\hline
\end{longtable}

\renewcommand*{\arraystretch}{1.3}
\begin{longtable}{llllllllll}
\caption{NEXA stellar parameters of our target list that were used in our study. Missing values are denoted with "--".}
\label{table:nexastarparams}
\\ \hline
       TIC ID &      Host Name &   Age & $\sigma_{\text{Age}}$ &   [Fe/H] & $\sigma_{\text{[Fe/H]}}$ & $T_{\text{eff,$\star$}}$ & $\sigma_{T_{\text{eff,$\star$}}}$ &   $K$ & $\sigma_{K}$ \\
              &                & [Gyr] &                 [Gyr] &          &                          &                      [K] &                               [K] & [mag] &        [mag] \\
\hline
TIC 298663873 &       TOI-2180 &  8.10 &                  1.40 &     0.25 &                     0.06 &                     5695 &                                59 &  7.60 &         0.02 \\
TIC 172370679 &       TOI-1899 &  7.10 &                  4.65 &     0.28 &                     0.11 &                     3926 &                                46 & 10.51 &         0.02 \\
TIC 377780790 &      Kepler-10 & 10.60 &                  1.40 &  $-$0.15 &                     0.04 &                     5708 &                                28 &  9.50 &         0.02 \\
TIC 432254760 & EPIC 249893012 &  9.00 &                  0.55 &     0.20 &                     0.05 &                     5430 &                                85 &  9.71 &         0.02 \\
 TIC 73717937 &        HATS-17 &  2.10 &                  1.30 &     0.30 &                     0.03 &                     5846 &                                78 & 10.70 &         0.02 \\
\hline
\end{longtable}

\section{\texttt{jaxstar} Parameters}\label{sec:jaxstarparams}

\renewcommand*{\arraystretch}{1.3}
\begin{longtable}{lllllllllll}
\caption{Homogenously-derived stellar parameters using \texttt{jaxstar} for the host stars of our sample.}
\label{table:jaxstarparams}
\\ \hline
       TIC ID &   Age & $\sigma_{\text{Age}}$ &   [Fe/H] & $\sigma_{\text{[Fe/H]}}$ &   $M_{\star}$ & $\sigma_{M_{\star}}$ &   $R_{\star}$ & $\sigma_{R_{\star}}$ & $T_{\text{eff,$\star$}}$ & $\sigma_{T_{\text{eff,$\star$}}}$ \\
              & [Gyr] &                 [Gyr] &          &                          & [$M_{\odot}$] &        [$M_{\odot}$] & [$R_{\odot}$] &        [$R_{\odot}$] &                      [K] &                               [K] \\
\hline
TIC 339672028 &  6.15 &                  1.28 &     0.27 &                     0.05 &          1.19 &                 0.06 &          1.65 &                 0.02 &                     5757 &                                67 \\
 TIC 94986319 &  8.79 &                  2.95 & $-$0.009 &                    0.048 &          0.85 &                 0.03 &         0.867 &                0.009 &                     5343 &                                59 \\
TIC 273231214 &  4.02 &                  0.56 &    0.023 &                    0.002 &          1.15 &                 0.02 &          1.35 &                 0.02 &                     6177 &                                58 \\
TIC 199376584 &  4.48 &                  1.39 &     0.04 &                     0.06 &          1.11 &                 0.05 &          1.26 &                 0.02 &                     6092 &                                97 \\
TIC 281885301 &  4.68 &                  3.27 &     0.05 &                     0.05 &          0.88 &                 0.03 &          0.83 &                 0.01 &                     5283 &                                73 \\
\hline
\end{longtable}

\twocolumn

\begin{figure}
  \centering
  \includegraphics[width=\linewidth]
  {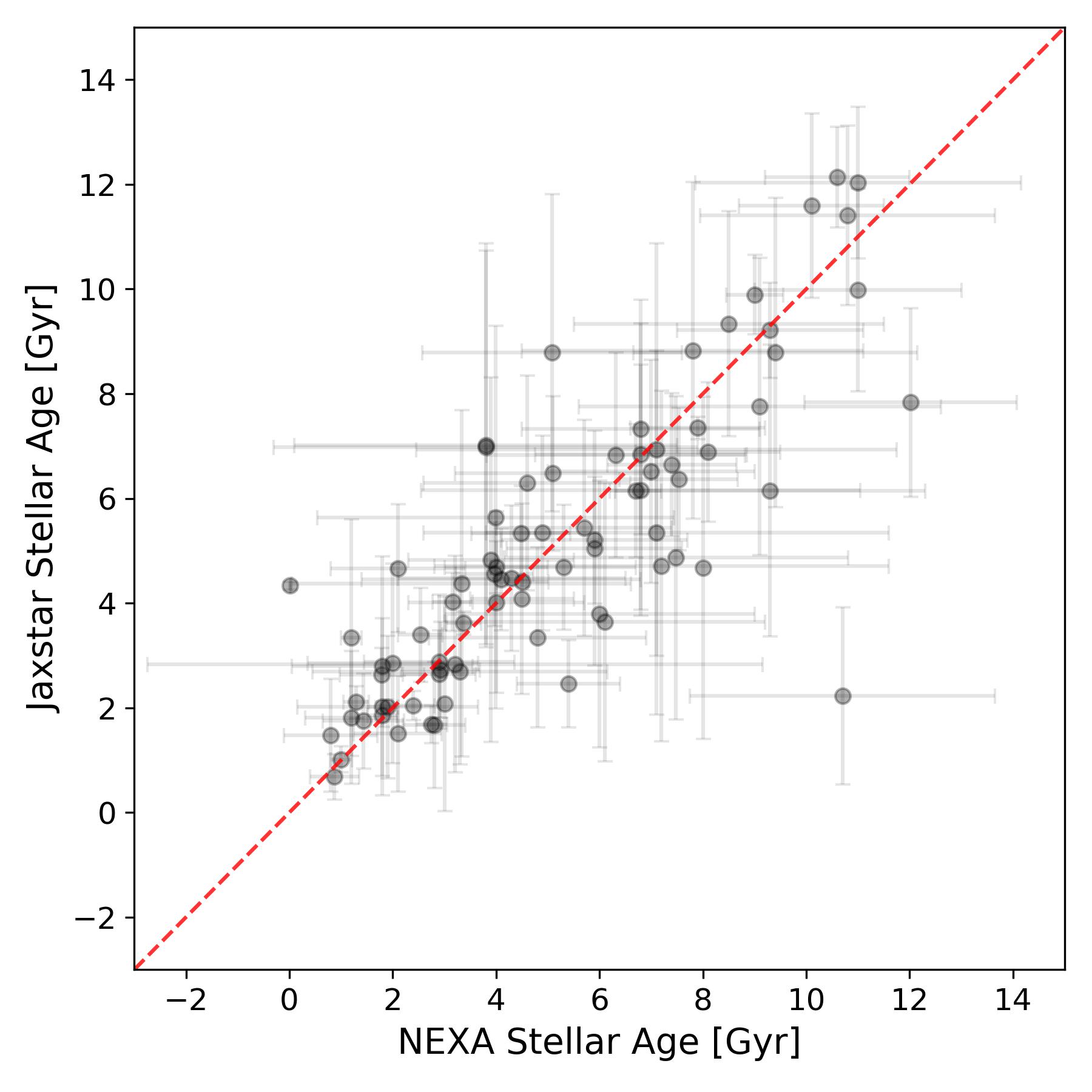} 
  \caption{NEXA vs. \texttt{jaxstar}-derived stellar ages.} 
  \label{fig:age_nexa_vs_jaxstar}
\end{figure}

\begin{figure}
  \centering
  \includegraphics[width=\linewidth]
  {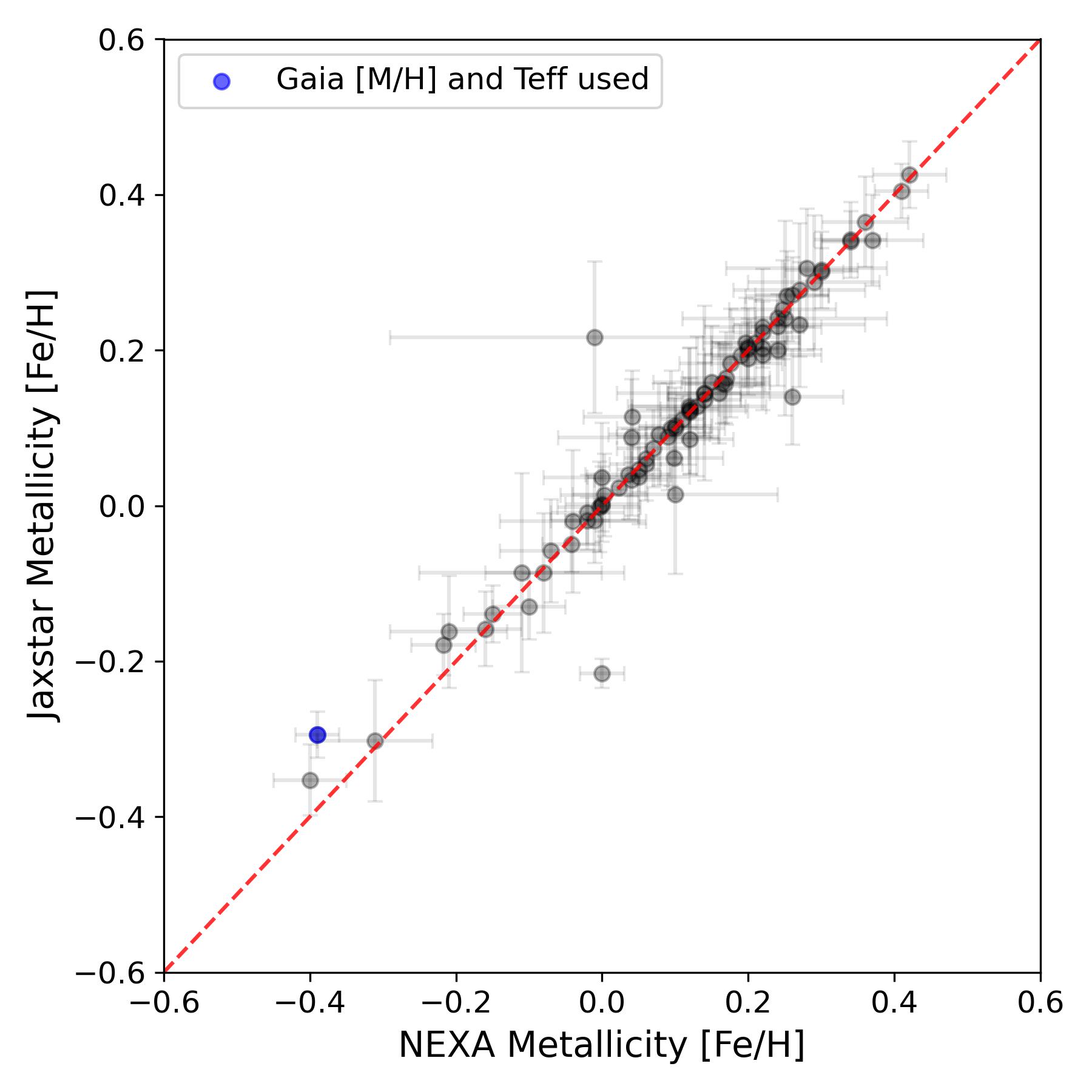} 
  \caption{NEXA vs. \texttt{jaxstar}-derived stellar metallicities.} 
  \label{fig:feh_nexa_vs_jaxstar}
\end{figure}

\begin{figure}
  \centering
  \includegraphics[width=\linewidth]
  {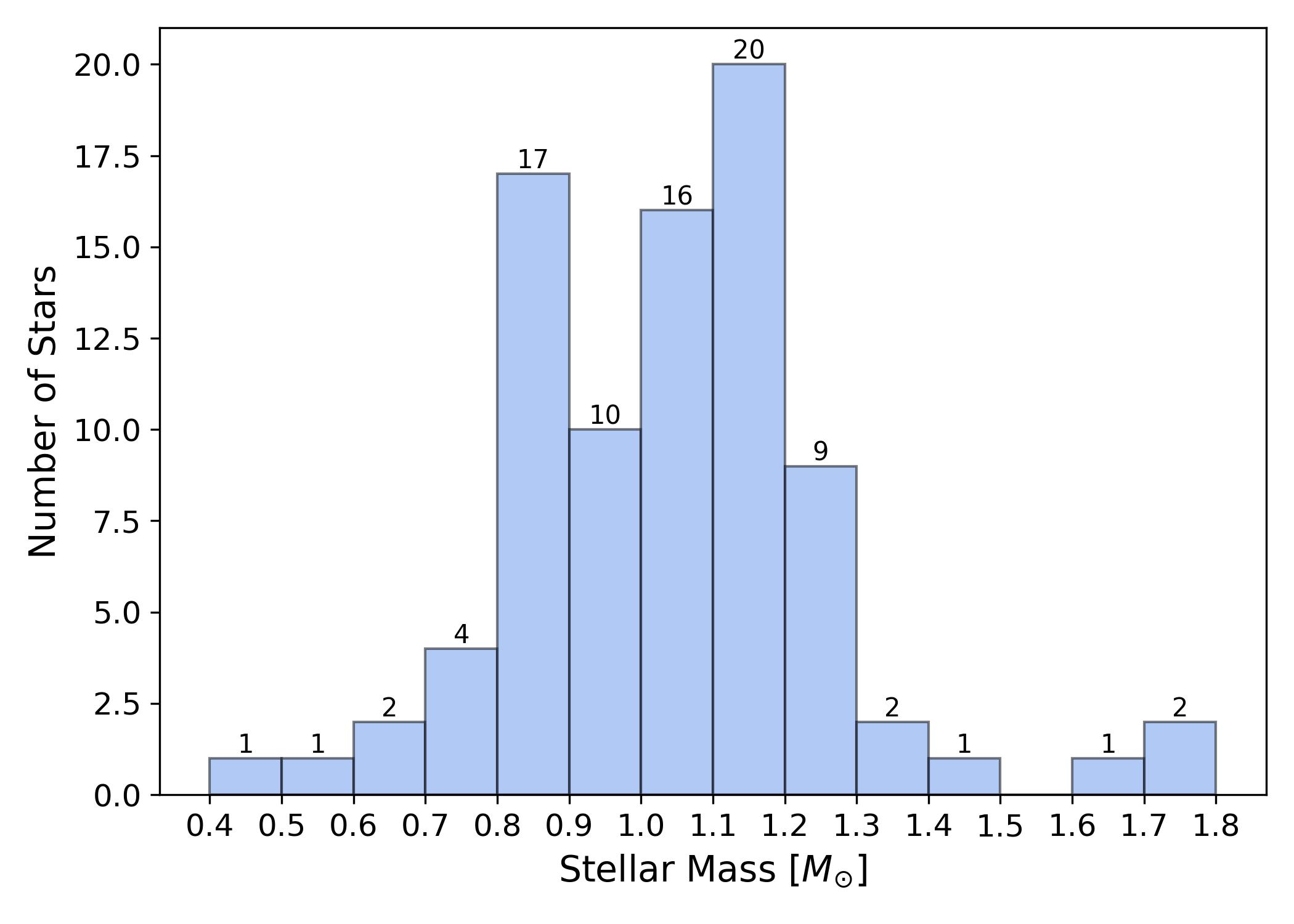} 
  \caption{Distribution of \texttt{jaxstar}-derived stellar masses for the host stars in our sample.} 
  \label{fig:jaxstar_mass_hist}
\end{figure}


\bsp	
\label{lastpage}
\end{document}